\documentclass[prc,twocolumn,floatfix,groupedaddress,nofootinbib,preprintnumbers,amsmath,amssymb,amsfonts,
superscriptaddress,widetable,href]{revtex4}
\usepackage{amsmath,amssymb,amsfonts}
\usepackage{hyperref}
\usepackage{ulem}
\usepackage{graphicx}
\usepackage{color}
\usepackage{longtable}

\newcommand{\be}{\begin{equation}} \newcommand{\ee}{\end{equation}}
\newcommand{\bea}{\begin{eqnarray}} \newcommand{\eea}{\end{eqnarray}}

\allowdisplaybreaks
\begin{document}

\title{\bf Generic size dependences of pairing in ultrasmall systems: electronic nano-devices and atomic nuclei}

\author{A. Pastore}
\affiliation{Department of Physics, University of York,\\
Heslington York YO10 5DD U.K.}
\author{P. Schuck}
\affiliation{Universit\'e Paris-Saclay, CNRS, IJCLab, IN2P3-CNRS, 91405 Orsay, France\\
Universit\'e Grenoble Alpes, CNRS, LPMMC, 38000 Grenoble, France}
\author{X. Vi\~nas}
\affiliation{Departament de F\'isica Qu\`antica i Astrof\'isica and Institut de Ci\`encies del Cosmos,
Facultat de F\'isica, Universitat de Barcelona, Diagonal 645, E-08028 Barcelona, Spain}

\date{\today}

\begin{abstract}
Average, global pairing behaviours of electronic devices, like films, wires, and grains are studied with semiclassical methods, 
as Weyl and Thomas-Fermi approximations,  in the ultrasmall, i.e., quantal regime, which exhibits strong quantum fluctuations 
and shell effects. This is the case also for superfluid nuclei. Results, mostly analytic, are elaborated for average size dependences, 
ready for easy use also in other circumstances. Comparisons with experimental results are given where possible.

\end{abstract}

\maketitle

\section{Introduction}


Atomic nuclei were the first objects where strong shell effects as a function of particle (nucleon) number, i.e. as a function of size, have been observed, see, e.g., \cite{RS, BM}. These structures have been explained since long using Hartree-Fock-Bogoliubov (HFB) or BCS theory \cite{RS, BM}. We will give examples in the main text. Pairing in metallic nano systems has been more intensively studied only in recent years with the advent of technological procedures to fabricate nano grains, films and wires. In those condensed matter systems one rather talks about the Bogoliubov de Gennes (BdG) approach to which reduces HFB for zero range pairing interactions. Also the 'diagonal' approximation of the gap is in use in nuclear physics as well as in condensed matter systems, in the former it is called BCS because of its similarity with the BCS equations of homogeneous matter whereas for condensed matter nano systems one usually talks about the 'Anderson' approximation because it was P.W. Anderson who proposed this in 1959 \cite{Anderson}. It turns out that the BCS/Anderson approximation works reasonably well both in nuclear and metallic systems. This stems from the qualitatively similarly shaped mean field in both cases with a flat interior and a steep surface, that is a Fermi function like potential. We will see in the main text why this is so. As just mentioned, in metallic nano systems great experimental achievements have been accomplished in the last 1-2 decades, but theory has not been at rest and impressive works for the numerical solution of nano-sized superconductivity have been published in recent years \cite{Shanenko06,Peeters,Shanenko07,Shanenko09,Shanenko11,G-G, Nature,GG1, GG2}. In these references also experimental results are given when they exist together with their citations.\\

In this paper, our emphasis is not so much on repeating or improving the exact solutions of the HFB/BdG equations, but our interest is mainly in the study of the average and generic size dependences of pairing in the different systems.  For nuclei this size dependence has been described by the purely empirical law $12/\sqrt{A}$ see, e.g., \cite{RS, BM}  where $A$ is the number of nucleons in the nucleus. This law describes the average trend reasonably well. It turns out that also in metallic nano-systems pairing becomes, on average, stronger with reduced sizes. It is one of the main objectives of this work to explain the underlying general reasons for this behaviour. The employed tools for this investigation shall be semi-classical methods, mainly based on Thomas-Fermi (TF) and Weyl approaches. Actually one of the earliest papers where the Weyl approach has bien applied to ultra-thin films where two of the present authors have been implied is given in \cite{grain}. There the dependence of the gap on the film thickness has been studied. In the present work we will make a more extended investigation and also consider other nano-devices as wires and grains as well as finite nuclei.\\

The paper is organised as follows. In Sec.\ref{sec:TF}, we  will explain  how one can successfully exploit TF and Weyl theories for the solution of the gap equation. In Sec. \ref{sec:BCS}  we will show that for 'steep' mean field potentials the BCS/Anderson approximation is valid to good accuracy. Then, in Sec.\ref{sec:metal}, we will demonstrate with generic examples like metallic films, wires, and grains their dependence on changing sizes. In Sec.\ref{sec:nuclei}, we will outline the situation concerning nuclei and we will present our conclusions in Sec.\ref{sec:conlcusion}.

\section{The Thomas-Fermi and Weyl Approaches to pairing in finite Fermi systems}\label{sec:TF}

For completeness, in this section, let us give a compact presentation of what was accomplished previously with a semiclassical approximation of what  one calls in nuclear physics the BCS method, a name which we also want to keep in this paper for brevity. This method assumes that the gap field $\Delta_{nn'}$ has the same form as in infinite homogeneous matter acquiring the following diagonal expression

\be\label{eq:delta}
\Delta_{nn'} \simeq \Delta_{n\bar n}\delta_{n',\bar n} \equiv \Delta_n\delta_{n',\bar n}\;,
\ee
where the single particle (s.p.) states, labeled by the index $n$ correspond to the mean field potential chosen for the problem at hand (e.g., the Hartree-Fock potential) and $\bar n$ is the time reversed state of $n$.
This approximation is valid, as we will show, for situations at equilibrium and at time reversal invariance with mean field potentials which have rather steep surfaces as they are encountered for real systems as, e.g.,  nuclei and metallic nano-particles. On the contrary, in cold atom systems, most of the time a wide harmonic oscillator confining external potential is used invalidating completely the use of the BCS approximation.
So let us assume that the BCS approximation is valid. The gap equation can be written as

\be
\Delta_{n} = - \sum_{n'}V_{nn'}\frac{\Delta_{n'}}{2E_{n'}}\;,
\ee
where $V_{nn'}=\langle n\bar n|v|n'\bar n'\rangle$ is the matrix element of the pairing interaction. In this paper, for simplicity, we will restrict ourselves to spin singlet pairing, which takes care of the antisymmetrisation of the matrix elements. So in the following of this work, we will consider always systems where spin singlet pairing is largely dominant and, thus, we can henceforth disregard the spin. Furthermore,

\begin{equation}
  E_n=\sqrt{(\epsilon_n-\mu)^2 + \Delta_n^2}\;,
  \label{qp-E}
  \end{equation}
  are the usual quasi-particle energies, with $\epsilon_n$ the single particle or Hartree-Fock (HF) energies of the mean-field Hamiltonian, that is $H^{\rm mf}|n\rangle = \epsilon_n|n\rangle$. The chemical potential $\mu$ is, as usual, determined by the particle number condition

\begin{equation}
  N = \sum_n\frac{1}{2}\bigg [ 1 - \frac{\epsilon_n - \mu}{E_n} \bigg ]\;.
  \label{N}
  \end{equation}

At equilibrium and for time reversal invariant systems, canonical conjugation and time reversal operation are related by $\langle {\bf r}|\bar n\rangle = \langle n|{\bf r}\rangle$. Consequently, we have $\langle {\bf r}_1{\bf r}_2|n\bar n\rangle = \langle {\bf r}_1|\hat \rho_n|{\bf r}_2\rangle$, with $\hat \rho_n = |n\rangle \langle n|$ the s.p. density matrix corresponding to the state $|n\rangle$. Therefore the pairing matrix element can be written as

\be
V_{nn'}=\langle n\bar n|v|n'\bar n'\rangle = \int d^3r\int d^3r'\rho_n({\bf r})
v({|\bf r}-{\bf r}'|)\rho_{n'}({\bf r}'),
\ee
with $v(r)$ the pairing force and $\rho_n({\bf r}) = \langle {\bf r}|n\rangle \langle n |{\bf r}\rangle$, the local density of the state $|n\rangle$.
The density matrix fulfils the Schr\"odinger equation

\be
(H-\epsilon_n)\hat \rho_n = 0,
\label{DM}
\ee
where $H$ is the Hamiltonian of the system; therefore, we can write $\Delta_n = Tr[\hat \Delta \hat \rho_n]$ and $\epsilon_n = Tr[H\hat \rho_n]$ and consequently the state dependence of the gap equation (1) is fully expressed by the density matrix $\hat \rho_n$.\\

\noindent
    {\it Fermi-function like mean field potentials}\\

Performing the Wigner transform (WT) of Eq.(\ref{DM}) \cite{RS} and taking into account that the WT of the product of two single particle operators $\hat A$ and $\hat B$ equals, to lowest order in $\hbar$, the c-number product of the corresponding WTs, i.e., $A({\bf R}, {\bf P})B({\bf R}, {\bf p})$, one easily obtains the $\hbar$-limit of Eq. (\ref{DM}). See Ref.\cite{RS} for more details.

\be
(H_{\rm cl.} - \epsilon)f_{\epsilon}({\bf R}, {\bf p}) = 0,
\label{sc}
\ee
where  $H_{\rm cl.} = \frac{p^2}{2m^*({\bf R})} + U({\bf R})$ is the classical Hamiltonian which contains a local mean field potential $U({\bf R})$ and a position dependent effective mass $m^*({\bf R})$ and $f_{\epsilon}({\bf R}, {\bf p})$ is the WT of $\hat \rho_n$. Equation (\ref{sc}) has to be read in the sense of distributions. Taking into account that $x\delta(x)=0$, where $\delta(x)$ is the Dirac delta function, here with $x= H_{\rm cl.} - \epsilon$,  one obtains for the normalised dist ribution function \cite{Faessler, Annals}

\be
f_{E}({\bf R}, {\bf p}) = \frac{1}{g^{\rm TF}(E)}\delta(E-H_{\rm cl.}) + O(\hbar^2),
\ee
which corresponds to the Thomas-Fermi (TF) approximation of the normalised on-shell or spectral density matrix \cite{50yearsBCS, grain}. Its norm is equal to the level density (without spin-isospin degeneracy)

\be
g^{\rm TF}(E) = \int \frac{d{\bf R}d{\bf p}}{(2\pi \hbar)^3}\delta(E-H_{\rm cl.}).
\ee
Corrections to those expressions in powers of $\hbar$ can be found in textbooks \cite{RS}. 
The semiclassical pairing matrix element can then be written as \cite{50yearsBCS, grain}

\begin{widetext}
\be
V(E,E') = \int \frac{d{\bf R}d{\bf p}}{(2\pi \hbar)^3} 
\int \frac{d{\bf R}'d{\bf p}'}{(2\pi \hbar)^3}f_{E}({\bf R}, {\bf p})
f_{E}({\bf R}', {\bf p}')v({\bf R}, {\bf p}; {\bf R}', {\bf p}'),
\ee
\end{widetext}
where $v({\bf R}, {\bf p}; {\bf R}', {\bf p}')$ is the double WT of $\langle {\bf r}_1{\bf r}_2|v|{\bf r}'_1{\bf r}'_2\rangle$. For a local translationally invariant force, this matrix element reduces to $v({\bf R}, {\bf p}; {\bf R}', {\bf p}') = \delta({\bf R}-{\bf R}')v({\bf p}-{\bf p}')$, with $v({\bf p}-{\bf p}')$ the Fourier transform of the force $v({\bf r}-{\bf r}')$ in coordinate space.\\
The gap equation in TF approximation is obtained by replacing in Eq.\ref{eq:delta}, $\hat \rho_n$ and $V_{nn'}$ by 
their corresponding semiclassical counterparts. In this way, the TF gap equation reads

\begin{figure*}
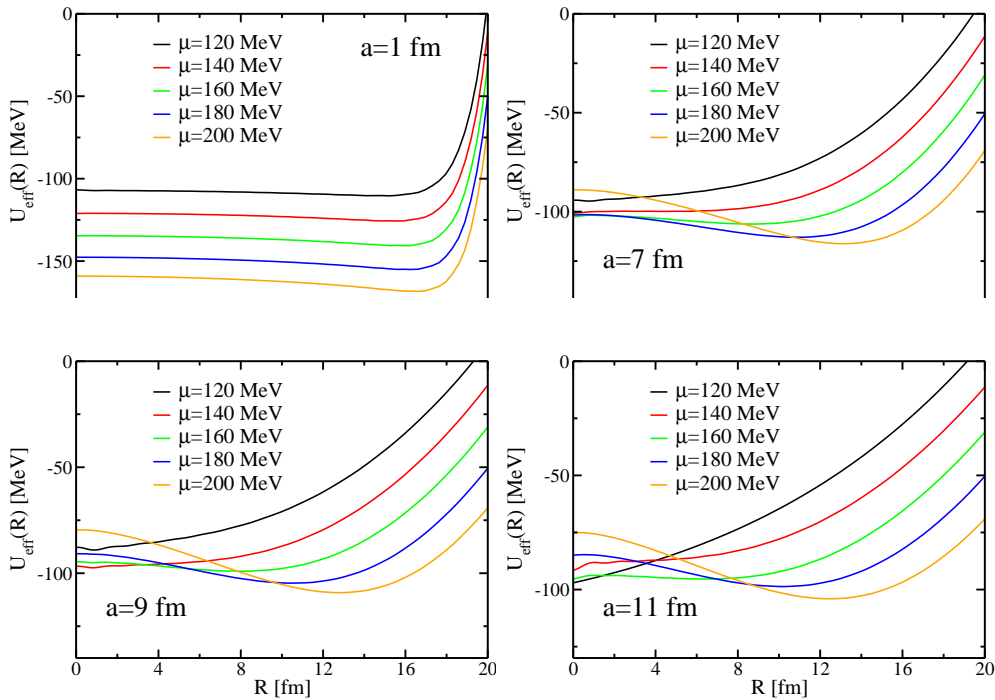

\includegraphics[width=6.5cm,angle=0]{pocket-WSa1.eps}
\includegraphics[width=6.5cm,angle=0]{pocket-WSa7.eps}
\\
\includegraphics[width=6.5cm,angle=0]{pocket-WSa9.eps}
\includegraphics[width=6.5cm,angle=0]{pocket-WSa11.eps}
\caption{\label{pockets}Pockets of different Woods-Saxon potentials with varying width parameters a = 1, 7, 9, 11 fm. }
\end{figure*}

\be
\Delta(E)= - \int_0^{\infty} dE'g^{\rm TF}(E')V(E, E')\kappa(E'),
\label{D(E)}
\ee
with

\be
\kappa(E) = \frac{\Delta(E)}{2E_{\rm q.p.}(E)}~;~~~E_{\rm q.p.}(E) = \sqrt{(E-\mu)^2 + \Delta^2(E)}.
\label{kappa(E)}
\ee
Eq.(\ref {D(E)}) can readily be solved for a given mean field potential and pairing force. The chemical potential $\mu$ is fixed by the  particle number condition (\ref{N})

\be
N= 2\int_0^{\infty} dEg^{\rm TF}(E)\frac{1}{2}\bigg [ 1-\frac{E-\mu}{\sqrt{(E-\mu)^2+\Delta^2(E)}}\bigg ],
\ee
where we introduced a factor of 2 for spin.\\
Mostly, we will use a zero range pairing force $-g\delta({\bf r}_1 - {\bf r}_2)$. In such a case one either introduces a cut-off or the gap equation (\ref{D(E)}) has to be regularised in the following way 

\begin{equation}
  \Delta(E) = -4\pi \frac{\hbar^2}{m}a\int dE' g^{\rm TF}(E') V(E,E')\bigg [\kappa(E') - \frac{\Delta(E')}{2|E'-\mu|}\bigg ],
\end{equation}
where $a$ is the corresponding scattering length \cite{Grasso}.\\
One may also be interested to transform the pairing tensor $\kappa(E)$ into coordinate space. In analogy to the quantum mechanical expression, this is given by

\be
\kappa({\bf R}) = \int dE g^{\rm TF}(E) \kappa(E)\rho_E({\bf R}),
\ee
with the local spectral density

\be
\rho_E({\bf R}) = \int \frac{d^3p}{(2\pi \hbar)^3}f_{E}({\bf R}, {\bf p}).
\ee\\

\noindent
    {\it Box potentials}\\

For potentials with infinite steep walls, e.g., the so-called box-potentials, the TF approximation can still be employed to 
lowest order what leads to the Fermi gas approximation. However, $\hbar$ correction terms as they are given by the Wigner Kirkwood 
expansion \cite{RS} cannot be used since the latter is based on smooth potentials which allow for a gradient expansion. 
In the case of box potentials the so-called Weyl expansion, further developed by Balian and Bloch \cite{Bal-Bloch} must be used. The WT of the on-shell density matrix is thus given by \cite{grain}

\begin{widetext}
\begin{equation}
f_{E}({\bf R},{\bf p}) =
\frac{1}{g(E)}\bigg[ \delta\left(E - \frac{\hbar^2 p^2}{2m}\right) - 
  \frac{2m}{\hbar^2} \delta p_z \frac{\cos(2 R k_{E}(p_x,p_y))}{k_{E}(p_x,p_y)} \bigg]\Theta(R_0(\theta,\varphi)- R(\theta,\varphi))
\label{Weyl},
\end{equation} 
\end{widetext}
where $R_0$ is the border of the box and $k_{E}(p_x,p_y)= \sqrt{\frac{2m}{\hbar^2}\big(E - \frac{\hbar^2}{2m}(p_x^2 + p_y^2)\big)}
= \sqrt{p_{E}^2 - p_{\perp}^2}$.
Performing the integral over momenta one obtains the spectral density 
\begin{equation}
\rho_E{\bf R}) = \frac{1}{4 \pi^2}\frac{2m}{\hbar^2}\frac{p_{E}}{g(E)} \big[ 1 - j_0(2 R p_ {E})\big],
\end{equation}
as it is explained in Appendix A. 
From the spectral density, one gets the density of states (DOS) in integrating over ${\bf R}$ as

\begin{equation}
  g(E) = \frac{1}{4\pi^2}\bigg (\frac{2m}{\hbar^2}\bigg )^{3/2}\sqrt{E}V -   \frac{2m}{\hbar^2}\frac{S}{16\pi},
\label{Weyl-DOS}
\end{equation}
where $V$ is the volume and $S$ the surface of the container. In principle, there is also a curvature term which, however, we will not consider here. In the above expressions, we should be aware of the fact that the origin of the $R$ coordinate is the surface and not the center of the system. $R$ is therefore to be considered as the linear distance variable from the surface and not as a radius. We, however, want to keep the notation common in the literature.\\
Next we will study under which conditions the BCS approximation can be employed.

\section{Validity of the BCS approximation. steep vs soft potentials}\label{sec:BCS}

In this section, we discuss the validity of the BCS approximation. To this end we investigate the behaviour of the effective potential 
  $U_{\rm eff}$, which is obtained from the elimination of the small component  of the 
BCS equations and taking the TF limit considering the cut of momenta along the local Fermi momentum 
$k_F(R)=\sqrt{\frac{2m}{\hbar^2}(\mu - U(R))}$. This effective potential reads

\begin{equation}
U_{\rm eff}(R) = U(R) - \mu + \frac{\Delta^2(R)}{E_{\rm qp}}.
\label{U-eff}
\end{equation}
In this equation $U(R)$ is the external mean field potential, $\Delta(R)$ is the quantal pairing field, and the quasi particle 
energy in local Fermi-momentum approximation is given by: 
$E_{qp} \simeq \sqrt{\left(\frac{p^2}{2m} - \mu(R)\right)^2 + \Delta^2(R)}$ where $\mu(R) = \frac{\hbar^2k_F^2(R)}{2m}$ 
is the local Fermi-energy.

In  Fig.\ref{pockets}, we take an example for nuclear dimensions, but the dimensions can be changed to electronic 
ones and the example has generic validity. We see that the effective potential $U_{\rm eff.}(R)$ forms a pocket at the 
surface for large values of $a$ parameter whereas it is absent for very steep potentials.
The potential and parameters used are:
$U(R)= -|U_0|[1- e^{(R-R_0)/a}]^{-1}$ with $|U_0| = 1200$ MeV ; $R_0 =22.12$ fm for $a=1$ fm; $R_0 = 34.84$ fm for $a=7$ fm; $R_0 = 39.08$ fm for $a=9$ fm; $R_0 = 43.32$ fm for $a=11$ fm. 
 Once there is a pronounced and wide pocket 
close to the surface, it is understandable that localised states can be formed in this pocket, see also \cite{Ale}, and 
that this particular feature cannot be reproduced by the BCS approach. On the other hand for steep potentials, the shape of 
$U_{\rm eff}(R)$ is very similar to the original one. The potential with the wide surface  is typically the case for cold atoms. 
Since the mean field is then very much altered by the presence of the gap potential, the pure mean field wave functions are no 
good solution any more and to account for this the gap has to become non diagonal in the mean-field basis as this is the case 
with the HFB approach. It should be noted, however, that the formation of the pocket also depends very much on the value of 
$\mu$. For small $\mu$ almost no pocket can be seen, neither in the steep nor in the soft potentials. That is why for small 
$\mu$ (and small gaps) the BCS approximation and, thus, also TF-BCS works quite well. Of course, the shape of the pocket is 
also a function of the gap value, that is, of the strength of the pairing force. So the lesson to be retained is that for steep 
mean field potentials, which is the case for most of the physical systems existing in nature, the BSC approximation is quite 
valid. We will consider in this work such systems.

\section{Metallic nano-devices}\label{sec:metal}

In this section, we will deal with superconducting nano-systems such as thin metallic films, nano-wires, and nano-grains. As mentioned in the introduction, we are interested in the global average behaviour of these systems in order to extract eventual generic features and, also, provide easy to use analytical formulas for the average size dependences as far as possible. Actually, as mentioned,  in an earlier publication, we have already considered thin films \cite{grain}, but here we want to extend our study in various directions making use of the semiclassical Weyl approximation for mean field potentials with hard walls what seems to be a valid approximation for metallic nano-systems.

\subsection{Films}

As usual, the decisive quantity for pairing is the Densities of States (DOS) of the various systems. 
So, first, we want to deduce the DOS how it enters the gap equation for films. To this end we make the approximation that the 
pairing matrix element can be replaced by its semiclassical expression $\tilde v_0$, which takes into account that the matrix 
element of the pairing force depends on the  size of the system, see below
Eq.(\ref{sc-me}). 

The gap equation reads 

\begin{equation}
  \Delta_{n,p_{\bot}} \!\!= \!\frac{\tilde v_0 \Theta_{n,e_{\bot}}}{2LS}\! \sum_{n'}\!\!
  \int_{-\infty}^{\infty}\!\frac{d^2p'_{\bot}}{(2\pi\hbar)^2}\frac{ \Theta_{n,e_{\bot}}\Delta_{n'e'_{\bot}}S}{\sqrt{(E_{n'}\!\! + e_{p'_{\bot}}\!\! -\mu)^2\! + \Delta_{n'e'_{\bot}\!\!\!}}},
    \label{slab-gap}
\end{equation}
where $S$ is the surface and

\begin{equation}
\Theta_{n,e_{\bot}} = \Theta(\omega_D - |E_n+e_{\bot}-\mu|).
\end{equation}
\noindent   In this equation $\omega_D$ is the Debye frequency, $e_{p_\bot}=\frac{p_{\bot}^2}{2m}$ is the kinetic energy 
in plane direction and $E_n$ are the quantized single particle  energies across the film of thickness $L$  defined as

\begin{equation}
E_n = \frac{\hbar^2}{2m} \bigg (\frac{\pi}{L}n\bigg )^2;~~ n= 1, 2, ... 
\end{equation}
We also can define $\Delta_{ne_{\bot}} = \Delta \Theta_{n,e_{\bot}} $.
%
This yields for the gap equation

\begin{equation}
  1=\frac{\tilde v_0}{L}\frac{1}{2\pi}\frac{2m}{\hbar^2}\sum_{n'}\frac{1}{2}\int_0^{\infty}de'_{\bot}\frac{\Theta_{n',e'_{\bot}}}{2\sqrt{(E_{n'} + e_{p'_{\bot}} -\mu)^2 + \Delta^2}}.
  \label{slab-gap2}
\end{equation}
Making a change of variables
\begin{equation}
 \xi = E_n + e_{\bot} -\mu, \nonumber
 \end{equation}
we obtain with $\Delta \ll \omega_D$ the standard weak coupling expression

\begin{eqnarray}
  1&=&\frac{\tilde v_0}{L}\frac{1}{4\pi}\frac{2m}{\hbar^2}\sum_n\Theta(\mu -E_n)\int_{-\omega_D}^{\omega_D}d\xi \frac{1}{2\sqrt{\xi^2 + \Delta^2}}\nonumber\\
  &\equiv& \tilde v_0g_F(\mu)\ln\left(\frac{2\omega_D}{\Delta}\right).
  \label{slab-DOS}
\end{eqnarray}

\noindent That is

\begin{equation}
  \Delta = 2\omega_D\exp \bigg [-\frac{1}{\tilde v_0g_F(\mu)}\bigg ].
  \label{gap-FW}
\end{equation}

The DOS per volume for a slab of thickness $L$ is given by (notice that the surface cancels out)

\begin{equation}
  g_F(\mu) = \frac{1}{L}\sum_n\int \frac{d^2p_{\bot}}{(2\pi \hbar)^2}\delta\left(\mu-E_n-\frac{p_{\bot}^2}{2m}\right).
  \label{g-slab}
  \end{equation}
  
\noindent Performing the integral leads to

\begin{equation}
  g_F(\mu) = \frac{1}{L}\sum_n \frac{1}{4\pi}\frac{2m}{\hbar^2}\Theta(\mu-E_n).
  \label{g-slab2}
\end{equation}
This is the same quantal level density which enters above gap equation (\ref{slab-DOS}).

The semiclassical DOS for films is given by the Weyl formula for three dimensional bodies with infinite walls and volume $V$ and surface $S$ of the
container, see (\ref{Weyl-DOS}) \cite{Bal-Bloch}.
\begin{equation}
  g^{\rm sc}(E)= \frac{1}{4\pi^2}\bigg (\frac{2m}{\hbar^2}\bigg )^{3/2}\sqrt{E}V -\frac{1}{16\pi}\frac{2m}{\hbar^2}S.
  \label{Ba-Blo}
\end{equation}

However, we should use the matter volume $V_M = V - (3\pi/8k_F^B)S_M$ which contains the right number of particles instead of $V$ 
as explained in Ref.\cite{Sto-Farine}. The DOS per volume is then given by
\begin{eqnarray}
  g^{\rm sc}(E)/V_M &=& \frac{1}{4\pi^2}\bigg (\frac{2m}{\hbar^2}\bigg )^{3/2}\sqrt{E}\nonumber\\
  &&\!\!\!\!\!\!\!\! \!\!+  \frac{2m}{\hbar^2}\frac{1}{16\pi}\bigg [ \bigg (\frac{2m}{\hbar^2}\bigg )^{1/2}
    \frac{3\sqrt{E}}{2k_F^B} -1 \bigg ]\frac{S_M}{V_M}.
  \label{dos-E}
\end{eqnarray}
We should be aware of the fact that for a film the ratio $S_M/V_M=2/L_M$ and that the following figures are traced with respect to $L$ which is the box limitation. We, therefore, give the matter values in terms of the box limitations

\begin{equation}
  \frac{S_M}{V_M} = \frac{S}{V}\bigg (1+\frac{3\pi}{8k_F^B}\frac{S}{V}\bigg ).
\end{equation}
It seems to be a second order effect, but for small box sizes and small Fermi momenta it can play a role, since the gap is exponentially dependent on the parameters. However, unless stated differently, we  will always use $L \simeq L_M$ in the following figures.

At the Fermi energy expression (\ref{dos-E}) yields

\begin{equation}
  g^{\rm sc}_{F}(\mu)/V_M =  
  \frac{1}{4\pi^2}\frac{2m}{\hbar^2}k_F\bigg (1+\frac{\pi}{8k_F}\frac{S_M}{V_M} + ...\bigg ).
  \label{Weyl-F}
\end{equation}



\begin{figure}
  \includegraphics[width=8cm]{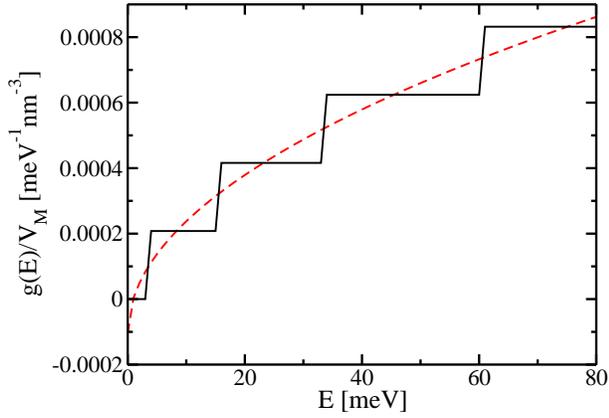}
  \caption{Comparison of quantal (solid) DOS with $L$ = 10 nm as a function of $E$ with the semiclassical expression (dashed), see  (\ref{dos-E}) (the constants for Cadmium \cite{Shanenko07} are used).}
  \label{DOS2L10}
\end{figure}

\noindent We compare in Fig.\ref{DOS2L10} the quantal (\ref{g-slab2}) (with $\mu$ changed to $E$) and semiclassical DOS (\ref{dos-E}) as a function of $E$ for $L$ = 10 nm. We see that we get a perfect average.

For the calculation of the gap, we give the semiclassical expression of the matrix element with its size dependence (
the derivation of this formula will be given in the Appendix 1)

\begin{equation}
  \tilde v_0 = v_0\bigg (1-\frac{\pi}{8k_F}\frac{S_M}{V_M}\bigg ) ,
  \label{sc-me}
\end{equation}
where $v_0$ is the bare coupling constant. This expression was confirmed in Ref. \cite{G-G}. We should be aware that for the calculation of the gap we need

\begin{equation}
  \tilde v_0 g^{\rm sc}_{F}(\mu) = v_0\frac{1}{4\pi^2}\frac{2m}{\hbar^2}k_F + O\left(\tfrac{S^2_M}{V^2_M}\right),
  \label{v0g}
  \end{equation}
and that therefore, to lowest order, a change of the product $\tilde v_0\tilde g_{F}(\mu)$ can only come from a size dependence of $k_F$. Indeed the semiclassical expression of the chemical potential is given by \cite{Sto-Farine}

 \begin{equation}
 \mu(L) = \mu_B\bigg (1+\frac{\pi}{2k_F^B}\frac{1}{L_M}\bigg ) ,
\label{musc}
 \end{equation}
 or

 \begin{equation}
   k_F(L) = k_F^B\bigg ( 1 + \frac{\pi}{4k^B_F}\frac{1}{L_M} \bigg ).
   \label{kFL}
   \end{equation}

 \noindent
 The chemical potential therefore rises as the size decreases. This the more, the smaller the system. This is due to the fact that the surface tension compresses the density and, thus, $\mu$ increases.
 
 Inserting Eq.(\ref{musc}) into the semiclassical DOS (\ref{Ba-Blo}) (with $E=\mu$) and the quantal one (\ref{g-slab2}), we can calculate both quantities as a function of film thickness $L$.

\begin{figure}
  \includegraphics[width=8cm]{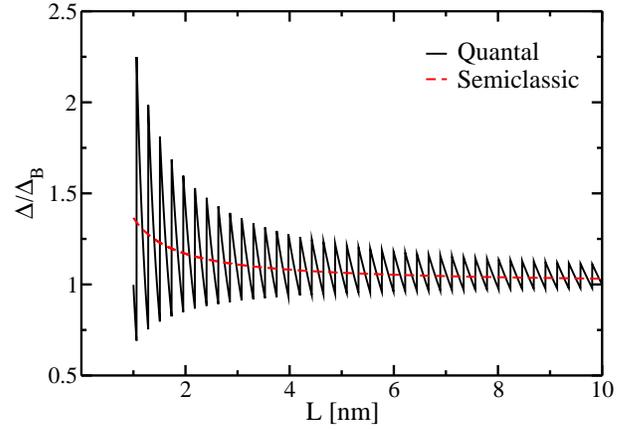}
  \caption{Quantal (solid) and semiclassical (dashed) normalised gap as a function of $L$ for Cadmium, see \cite{Shanenko07} for the constants.}
  \label{gap-qu-sc}
\end{figure}
\begin{figure*}
  \includegraphics[width=11cm]{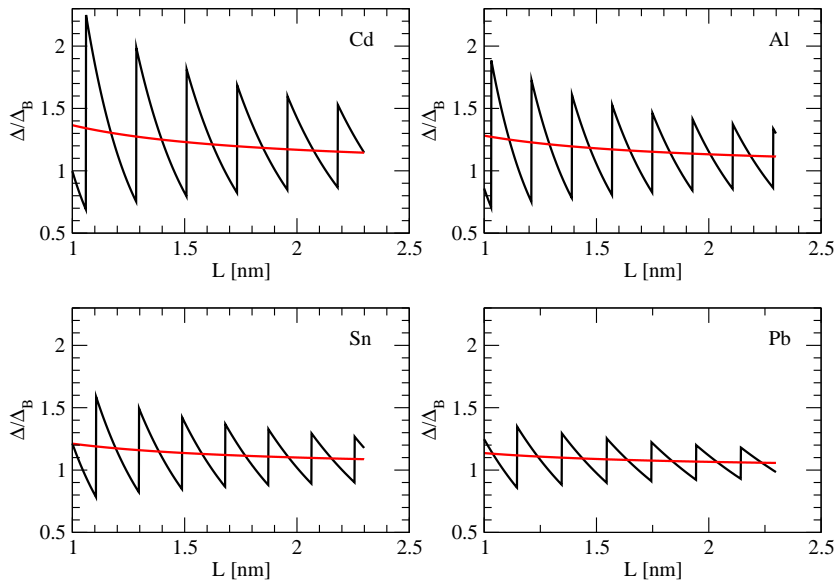}
  \caption{Quantal and semiclassical gap using FW for Cadmium (upper left), Aluminium (upper right), Tin (lower left), lead (lower right).
  We see almost perfect agreement with quantal solution of Shanenko in Fig.1 of \cite{Shanenko07}.}
  \label{gap-qu-sc2}
  \end{figure*}
Let us write out the semiclassical gap more explicitly.
We have to consider that the Fermi momentum increases with decreasing $L$ according to Eq.(\ref{kFL}).
Therefore the correction factor in the product $\tilde v_0 g^{\rm sc}_F(\mu)$  even for $L$=1 nm is only a correction of 4-5\% for Cadmium with the parameters given in \cite{Shanenko07}.

With Eq.(\ref{v0g}), this then leads to the semiclassical correction formula for the gap already obtained in \cite{grain}



\begin{equation}
  \frac{\Delta}{\Delta_B} = \exp\bigg [ -\frac{1}{|v_0|g_F^B(1+\frac{\pi}{4k_F^B}\frac{1}{L_M})}\bigg ]\exp\bigg (\frac{1}{|v_0|g_F^B}\bigg ).
    \label{Dwc-sc}
\end{equation}
We can also try to replace in this formula the semiclassical DOS by the quantal one. Since the latter does not deviate violently from the mean value, this may give reasonable results. For this, we have to be aware of the fact that the quantal DOS takes $V$ and not $V_M$ and, therefore, we have in the semiclassical limit $g(\mu)/g_F^B= 1-\frac{\pi}{4k_F^B}\frac{S}{V}$. 
In order to be consistent, we then have also to use for $\tilde v_0 = v_0(1+\frac{\pi}{4k_F^B}\frac{S}{V})$ 
which is the expression using the box quantities (see Appendix 1). However, here it makes numerically not much difference to replace $S/V$ by $S_M/V_M$. Together with the compression effect for the Fermi momentum, we obtain the following formula for the gap using the quantal DOS

\begin{equation}
  \frac{\Delta}{\Delta_B} = \exp\bigg [ -\frac{1}{|v_0|g_F^B(1+\frac{\pi}{2k_F^B}\frac{1}{L_M})\frac{g_F(\mu)}{g_F^B}}\bigg ]\exp\bigg (\frac{1}{|v_0|g_F^B}\bigg ),
    \label{Dwc-qu}
\end{equation}
where $g_F(\mu)$ is the quantal DOS given in (\ref{g-slab2})). For numerical values of Cd, Al, Sn, Pb, we take the ones given in \cite{Shanenko07}. In Fig.\ref{gap-qu-sc}, we show the gap, normalized to its bulk value for Cd for up to large values of $L$ in order to demonstrate the correct behavior of the semiclassical values versus the quantal ones.
In Fig.\ref{gap-qu-sc2} we show the results for Cd (upper left), Al (upper right), Sn (lower left), and Pb (lower right) using formulas (\ref{Dwc-sc}) and (\ref{Dwc-qu}). We see that, on the one hand, the fully quantal results of Fig.1 in \cite{Shanenko07} are well reproduced and, on the other hand, the semiclassical formula (\ref{Dwc-sc}) averages very well the quantal results (smooth continuous lines) in all cases.
\\Therefore, for films the average size behaviour is well given by the semiclassical analytic expressions (\ref{Dwc-sc}) and (\ref{Dwc-qu}). In conclusion, it is quite clear that the increase of the average gap with decreasing film thickness comes from the fact that the surface tension compresses the system more and more as its size decreases and, thus,$k_F$, i.e., the DOS increases. The formulas can be applied to other systems, if needed. We next will turn to the case of nano-wires.

\subsection{Wires}

Recently nice experiments have been performed on superconducting metallic  nano-wires as shown and cited in Ref.\cite{Peeters}. Those systems show shell effects which are much stronger than the ones of films. We again want to investigate the average pairing behaviour as a function of the wire radius.

Let us start with an infinitely long wire with diameter $2R$. A crucial ingredient is again the level density

\begin{equation}
  g_W(E) = \lim_{L\rightarrow \infty}\sum_iL\int \frac{dp_z}{2\pi \hbar}\delta(E-E_i-e_z),
  \label{wireLD}
\end{equation}
with $e_z = p_z^2/(2m)$ is the kinetic energy in wire direction,  and

\begin{equation}
E_i =\frac{\hbar^2}{2m}\frac{\alpha_{jm}^2}{R^2},
\label{disk-E-i}
\end{equation}
the eigenvalues of the transverse disk with radius $R$ and $\alpha_{jm}$ the zeros of the Bessel functions of the first kind, see \cite{Abramo-Steg}. For later convenience, let us write (\ref{wireLD}) also in the following way

\begin{equation}
  g_W(E) = \int d\omega g_D(\omega)g_z(E-\omega),
  \label{fold}
  \end{equation}
where $g_D(E)$ is the discrete level density of the disk

\begin{equation}
  g_D(E) = \sum_i\delta(E-E_i),
\end{equation}
and $g_z(E)$ the one in wire direction and given in Eq.\ref{1D-LD} below. The integral over $p_z$ can be performed and we obtain
\begin{widetext}
\begin{equation}
  g_{W,R}(E) = \frac{1}{2\pi^2}\frac{1}{R^2}\bigg ( \frac{2m}{\hbar^2} \bigg )^{1/2}\sum_{jm} \bigg [\delta_{m,0}\frac{\Theta(E - \hbar^2\alpha_{jm}^2/2mR^2)}{\sqrt{E - \hbar^2\alpha_{jm}^2/2mR^2}} + 2(1-\delta_{m,0})
    \frac{\Theta(E - \hbar^2\alpha_{jm}^2/2mR^2)}{\sqrt{E - \hbar^2\alpha_{jm}^2/2mR^2}} \bigg ].
  \label{wireLD2}
  \end{equation}
\end{widetext}
We will consider the DOS at $E=\mu$ as a function of disk radius. Then as a function of decreasing $R$, one level after the other will be resonant, i.e., pass past $\mu$. This gives a typical resonance structure in the gap.\\
  In reality $\mu$ also increases while $R$ becomes smaller, but this is a small effect: on average $\mu$, i.e., $k_F$ increases only a little while $R$ is decreasing.The semi-classical  expression for $\mu$ is with $S_M/V_M = 2/R_M$ given by

  \begin{equation}
 \mu(R) = \mu_B\bigg (1+\frac{\pi}{2k_F^B}\frac{1}{R_M}\bigg ).
\label{mu-sc}
  \end{equation}
  This $R$ dependence of the chemical potential averages very well the quantal behaviour.
Here and in the following we use the parameters given in Ref.\cite{Peeters} for Aluminium: 
$v_0N(0) = 0.18;~~ \omega_D = 32.3$ meV;~~$\mu_B = 900$ meV;~~the corresponding bulk Fermi momentum is then: 
$k_F^B = 4.83$ nm$^{-1}$; the bulk value of the gap: $\Delta_B = 0.25$ meV.

\begin{figure}
    \includegraphics[width=6.5cm, angle=0]{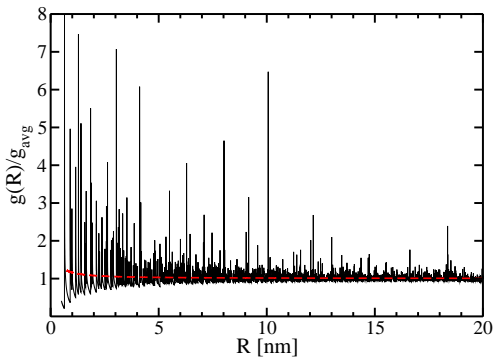}
    \caption{Quantal (solid) and semiclassical (dashed) DOS of wire at $E=\mu$ as a function of wire thickness.}
    \label{DOS-wire}
    \end{figure}

\begin{figure}
    \includegraphics[width=6.5cm, angle=0]{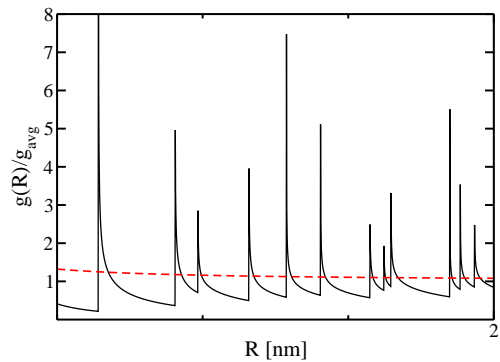}
    \caption{Same as Fig.\ref{DOS-wire}, but for very small sizes.}
    \label{DOS-wire2}
    \end{figure}

  
  The quantal level density of the wire is shown for $E=\mu(R)$ for Al as a function of wire thickness in Fig.\ref{DOS-wire} 
for $R< 20$ nm and in Fig.\ref{DOS-wire2} for $R~< ~ 2$ nm.\\

\begin{figure}
     \includegraphics[width=6.5cm,angle=0]{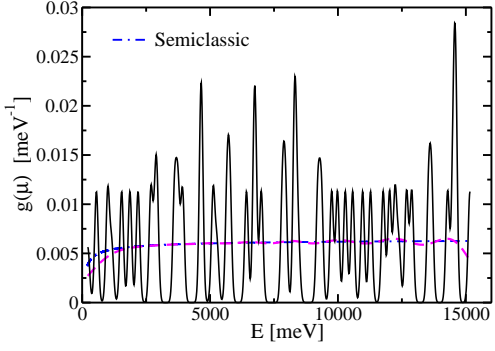}
     \caption{DOS of disk with radius $R=1$ nm averaged with a Gaussian of various width parameters $b$: solid: $b$=100 meV; dashed: 
$b$=1000 meV; dot-dashed: semiclassical values. We see that the  dashed curve well agrees with the semiclassical one, besides at 
the origin where the boundary distorts the Strutinsky average.}
     \label{disk-LD}
   \end{figure}

  From Fig.\ref{DOS-wire2}, we see that the semiclassical DOS again averages the quantal one very well. The discrete levels of the disk 
give rise to square root singularities embedded in the continuum of the wire DOS. We notice that the average DOS raises only very little 
towards very small diameters and it is practically flat elsewhere. This is due to the fact that the average DOS of a disk is a constant 
as we will explain now.
  Physical insight can again be gained from  the average level density of the wire. The average level density $\tilde g_D(E)$ of a disk 
  is known, see the book by Bhaduri and Brack\cite{Bhaduri-Brack},

  \begin{equation}
    \tilde g_D(E) = \frac{1}{4\pi}\frac{2m}{\hbar^2}\pi R^2 -\frac{1}{8\pi}\bigg (\frac{2m}{\hbar^2}\bigg )^{1/2}\frac{1}{\sqrt{E}}2\pi R.
    \label{av-g-D}
  \end{equation}
 In Fig.\ref{disk-LD}, we show the average 2D-disk DOS for a given $R$-value as a function of energy $E$. On the one hand we give the semiclassical DOS and on the other hand we also performed an averaging in smoothing the quantal DOS with a normalized Gaussian of increasing width $b$. We see that for $b=1000$ meV there is perfect agreement with the semiclassical result. Only at the borders there is some disagreement because of the boundaries with the Gaussian smoothing. In those regions we better trust the semi-classical results.
  The wire DOS can be written as a folding integral, as in (\ref{fold}), of the disk DOS with the one of homogeneous infinite matter in $z$-direction

  \begin{equation}
    g_z(E)=   L\int \frac{dp_z}{2\pi \hbar}\delta(E - e_z) = \frac{L}{2\pi}\bigg (\frac{2m}{\hbar^2}\bigg )^{1/2}\frac{1}{\sqrt{E}}.
\label{1D-LD}
\end{equation}
    
   With (\ref{av-g-D}), we  perform the folding integral to get the average DOS for the wire

   \begin{equation}
     \sum_i\int d\omega \tilde g_D(\omega - E_i)g_z(E-\omega),
     \label{folding}
   \end{equation}
   and obtain with

   \[ \int_0^Ed\omega \sqrt{\frac{1}{E-\omega}}= 2\sqrt{E}~;~~~
   \int_0^Ed\omega \sqrt{\frac{1}{\omega(E-\omega)}}= \pi . \]
The average wire DOS $ g_W^{\rm sc}$ is then given as
   
   \begin{equation}
     g_W^{\rm sc}(E)/V = \frac{1}{4\pi^2}\bigg (\frac{2m}{\hbar^2}\bigg )^{3/2}\sqrt{E} - \frac{2m}{\hbar^2}\frac{1}{16\pi}\frac{S}{V}
     \label{DOSsc-W},
   \end{equation}
   with $V=\pi R^2L$ and $S=2\pi RL$ for large values of $L$. This expression is exactly the same as obtained from the 3D level density 
with a finite size correction, Eq.(\ref{Weyl-DOS}).

  \begin{figure}
    \includegraphics[width=7.5cm,angle=0]{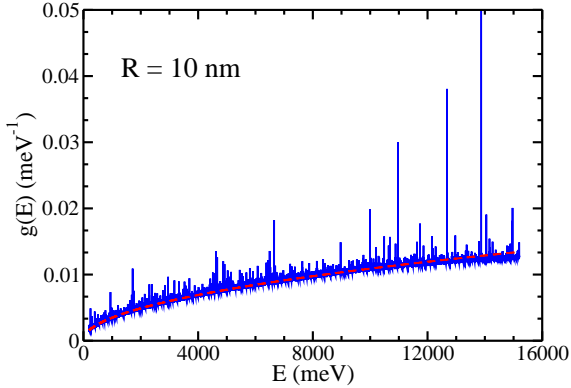}
    \caption{DOS for $R$= 10 nm as a function of energy}
    \label{DOSE-wire}
  \end{figure}
The expression involving the matter quantities $V_M$ and $S_M$ has already been given in eq.(\ref{dos-E}). 


It is instructive to show the performance of the semi-classical DOS as a function of energy for a given radius (here $R$=10 nm). 
This is shown in Fig.\ref{DOSE-wire} with the dashed curve. We notice again a very good representation of the average.
  In this figure we can see that from $ E \sim 12000 $ meV on, the quantal sampling becomes slightly deficient, since its average is undershooting the semiclassical curve. \\

\begin{figure}
  \includegraphics[width=8cm]{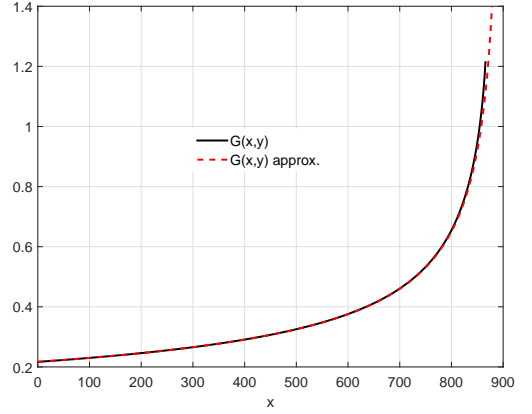}
  \caption{\label{elliptic} Approximate (dashed) and exact (solid) $e_z$ integral.
  Aluminium constants are taken ($\mu$ = 900 meV and $\omega_D$ = 32.3 meV). We see that in the range of $E$ between zero and 900-32.3 meV, that is the point where the black curve stops, the approximate solution is excellent. The exact function (solid) stops because from there on solution becomes imaginary.}
\end{figure}

  Knowing the quantal and average level densities, we can try, as in the case of films,  to recover the corresponding gap-values as a function of $R$ via the weak coupling solution of the gap equation. Supposing further, as for the film, that matrix elements are constant in this energy range because of the narrow Debye window  and that they are  equal to the semiclassical expression $\tilde v_0$ of (\ref{sc-me}), 
which nevertheless keeps a size dependence, we have (dividing (\ref{sc-me}) by the volume)

  \begin{equation}
 \!   \Delta_{i,p_z}\! = \!\frac{\tilde v_0}{ R^2L}\Theta_{i,p_z}\!\!\sum_{i'}L\!\!\int_{-\infty}^{\infty}\!\!\frac{dp'_z}{4\pi^2 \hbar}\frac{\Theta_{i',p'_z}\Delta_{i',p'_z}}{\sqrt{(E_{i'}\!+e_z\!-\!\mu)^2+\Delta_{i'\!,p'_z}^2}}.
    \label{wire-gap}
  \end{equation}
  We have used 
  
  \begin{eqnarray*}
\Theta_{i,p_z} &=& \Theta(\omega_D-|E_i + e_z -\mu|),\\
\Delta_{i,p_z} &=& \Delta \Theta_{i,p_z} .
  \end{eqnarray*}
  
\noindent   This yields for the gap equation

  \begin{equation}
    1=\frac{\tilde v_0}{\pi R^2}\frac{1}{2\pi}\bigg (\frac{2m}{\hbar^2}\bigg )^{\tfrac{1}{2}}
    \sum_i\int_0^{\infty}\frac{de_z}{\sqrt{e_z}}\frac{\Theta_{i,p_z}}
        {\sqrt{(E_{i}+e_z-\mu)^2+\Delta^2}}.
  \end{equation}
  The integral can be written as
  \begin{equation}
   1\!=\!\frac{\tilde v_0}{\pi R^2}\frac{1}{2\pi}\bigg (\frac{2m}{\hbar^2}\bigg )^{\tfrac{1}{2}}\!\!\!\!
   \int \!dE \!\int_0^{\infty}\!\!\frac{de_z}{\sqrt{e_z}} \sum_i\frac{\delta(E-\varepsilon_{i,z}) \Theta_{i,p_z}}
        {\sqrt{E^2+\Delta^2}},
   \label{eq51}
  \end{equation}  

\noindent where we have introduced the shorthand notation $\varepsilon_{i,z}=E_i+e_z-\mu$.
  As usual in weak coupling, we take the level density out of integral at peak value of the integrand, that is at $E=0$, and make a change of integration variable to

  \begin{figure*}
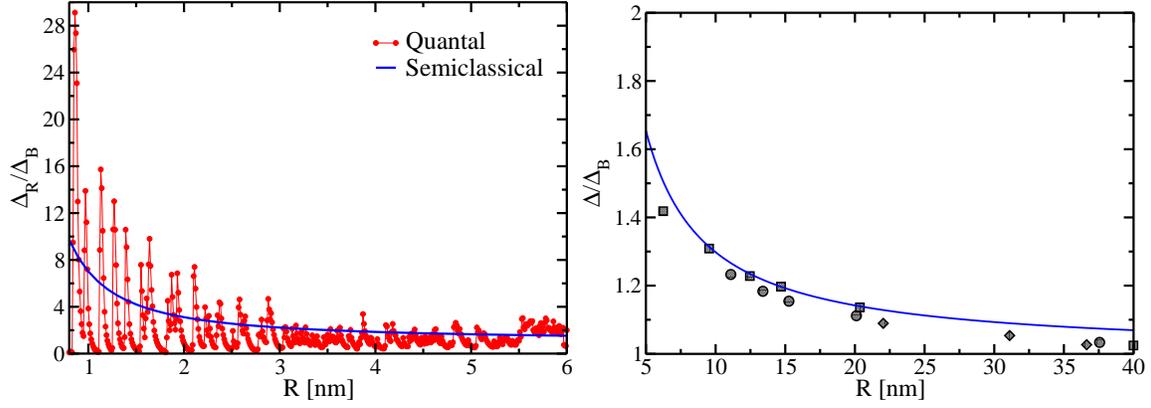

    \includegraphics[width=7.5cm]{Ale-wire1.eps}
    \includegraphics[width=7.5cm]{Ale-wire2.eps}
    \caption{\label{delta-B}Semiclassical and quantal results for Aluminium wires for different ranges of the wire diameter $D=2R$. In the left panel the quantal solution is compared with the semiclassical one (smooth continuous line). On the right panel, 
we remark the very good agreement of the semiclassical result with the experimental data \cite{Peeters}. Please look up Ref.\cite{Peeters} for the experimental data references because they are partially from private communication and, thus, cannot be given here.}
  \end{figure*}

 \begin{equation}
 \xi = E_i + e_z -\mu. \nonumber
 \end{equation}
  We then obtain

\begin{equation}
    1=\frac{\tilde v_0}{\pi R^2}\frac{1}{2\pi}\bigg (\frac{2m}{\hbar^2}\bigg )^{1/2}
    \sum_i\frac{1}{\sqrt{\mu - E_i}}\int_0^{\omega_D}d\xi \frac{1}{\sqrt{\xi^2 + \Delta^2}},
    \label{approxi-gap}
\end{equation}
Actually, the integral over $e_z$  in (\ref{eq51}) can be done without approximation, but the expression is in terms of the 
incomplete elliptic integral of the first kind which is not easy to handle. Comparing the exact solution with the approximation, one 
notices that the approximation is excellent besides very close to the divergency, as it can be seen in Fig.\ref{elliptic}

We recognise in (\ref{approxi-gap}) the wire DOS (\ref{wireLD2}).  Therefore we can use  
for the wire DOS appearing in (\ref{approxi-gap}) either its 
quantal (\ref{wireLD2}) or its semiclassical (\ref{DOSsc-W}) forms.






Performing the $\xi$ integral,
we then get the following equation for the semiclassical gap which has the same structure as the one for the films

\begin{equation}
  1=\tilde v_0g_W(R)\ln \frac{2\omega_D}{\Delta},
  \label{wc-wire-gap}
\end{equation}
or

\begin{equation}
  \Delta(R) = 2\omega_D \exp\left(-\frac{1}{\tilde v_0 g_W(R)}\right).
  \label{Dsc-W}
\end{equation}
Within the semiclassical approximation, we have

\begin{equation}
\tilde v_0 = |v_0|\left(1 - \frac{\pi}{4k_F}\frac{S_M}{V_M}\right), \nonumber
\end{equation}
and, thus with the size dependence of $k_F$, i.e., the compression effect

\begin{equation}
  \tilde v_0g_W(R) = v_0g_F^B\left(1+\frac{\pi}{4k_F^B}\frac{1}{R_M}\right).
  \label{sc-wc}
\end{equation}
See  in this respect (\ref{sc-me}, \ref{v0g}, \ref{kFL}) for the analogous expression for the case of films.\\

However, the expressions (\ref{Dsc-W}) with (\ref{sc-wc}) give, contrary to the case of the films, a too small gap with respect to the quantal values. This stems from the fact that the quantal DOS fluctuates in the case of the wire much more strongly than in the case of films, see Fig.\ref{DOS-wire2}. The DOS is an input to the gap equation and even though the semiclassical DOS perfectly averages the quantal DOS, it is not guaranteed that the solution of the gap equation using the semiclassical DOS as input also averages the fully quantal gap-solution well. In the case of the wires this is not the case, in contrast to the very satisfying solutions in the case of films. 
The semiclassical gap calculated with (\ref{sc-wc}) is much too small. This comes from the fact that the DOS for the wire is very asymmetric around the average, as can be seen in Fig.\ref{DOS-wire2}. The square-root infinities in the DOSenhance the gap very much what is not the case using the averqage DOS in the gap equation.  To compensate for this effect, we multiply Eq.(\ref{sc-wc}) by an  empirical enhancement factor

\begin{equation}
  \tilde v_0g_W(R) \rightarrow  v_0g_F^B\left(1+\frac{\pi}{4k_F^B}\frac{1}{R}\right)a(R)
\end{equation}
which can be found to be given very accurately by ($k_F^B$ = 4.83 nm$^{-1}$ for Al)

\begin{equation}
 a(R) =\bigg [\frac{g^{sc}_W(\mu)}{g_F^B}\bigg ]^2 \simeq 1+ \frac{\pi}{2k_F^B}\frac{1}{R} \simeq 1+0.325\frac{1}{R}.\nonumber
 \end{equation}
 
The corresponding semi-empirical average size-dependence of the gap

\begin{equation}
\frac{\Delta(R)}{\Delta_B} =\exp\bigg (-\frac{1}{|v_0|g_F^B(1+\frac{\pi}{4k^B_F}\frac{1}{R})a(R)}\bigg )\exp \bigg (\frac{1}{|v_0|g_F^B}\bigg ),
    \label{gapFWqu}
    \end{equation}
works very well (with $\pi/(4k_B^F) \simeq 0.162$). In Fig.\ref{delta-B} we show the semiclassical gaps as a function of $R$ up to 6 nm (left panel) and in the right panel the comparison with experiments. We see that in both cases our semiclassical formula (\ref{gapFWqu}) works very well.

For the approximate quantal gap normalized to the bulk, one could try to use  similarly to the case of films an expression analogous to (\ref{Dwc-qu}).
However, due to the already mentioned strongly asymmetrically fluctuating DOS of the wire with respect to its mean value, this formula is, contrary to the film case, inaccurate and yields even qualitatively unacceptable results.
If one nevertheless does this what means replacing $a(R)$ by $g_{\
  W,R}(\mu(R))/g_B$, and then smoothens in folding the result with a Gaussian, one\
gets back rather closely the average full line in Fig.\ref{delta-B}. This demonstrates well that taking in the gap equation average quantities like the semiclassical DOS does not necessarily mean that the result also averages the quantal gap.

\subsection{Metallic grains}

\begin{figure}
  \includegraphics[width=7.5cm,angle=0]{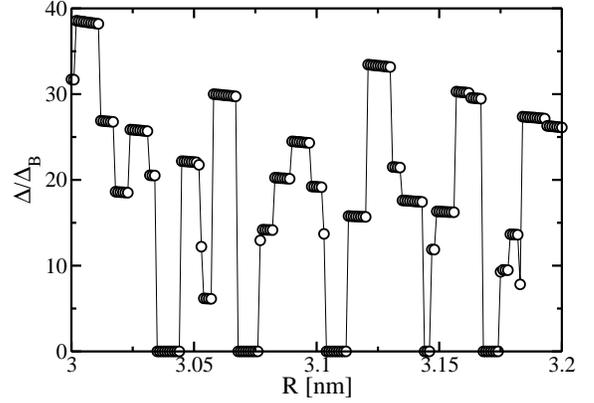}
  \caption{Strong zoom for gap of Sn of spherical grain. The plateau's correspond to single levels entering and leaving the Debye window.}
  \label{gapzoome}
\end{figure}

For the description of small metallic grains the box potential may again  be a valid model. However, there exists a quite strong particularity in the fact that all s.p. energies are discrete and  that the range of the force is limited by the Debye window $\mu \pm \hbar \omega_{\rm D}$ where $\omega_{\rm D}$ is again the Debye frequency. The gap equation then reads, see Ref.\cite{Shanenko11} for more details,
    
\begin{equation}
  \Delta_{\nu} = -\sum_{\nu' = \mu - \hbar \omega}^{\mu + \hbar \omega}\bar v_{\nu \bar \nu \nu' \bar \nu'}\frac{\Delta_{\nu'}}{2\sqrt{(e_{\nu'}-\mu)^2 +\Delta_{\nu'}^2}},
  \label{BCS-gap}
\end{equation}
with $e_{\nu}$ and $\bar v_{\nu \bar \nu \nu' \bar \nu'}$ are the single particle energies and the usual antisymmetrized matrix elements of the pairing force \cite{RS}.
The point now is that the Debye window is so small that for grains of few nanometers radius only very few discrete s.p. levels enter into the window. The situation is well described by Croitoru {\it et. al.} in \cite{Shanenko11}. So for box radii $R < 10$ nm practically only diagonal matrix elements of the force enter. In such a situation, it is very important to label the states by their individual quantum numbers $\nu=nlms$ where, in the order, they are principal quantum number, angular momentum, spin and angular momentum projection. Since we consider spin singlet pairing, we can forget about spin and have for the quantum numbers only $nlm$. In this case the above gap equation reads

\begin{equation}
  \Delta_{nl}= -(2l+1)\!\!\!\!\sum_{n'l' = \mu - \hbar \omega}^{\mu + \bar \omega}v_{nln'l'}\frac{\Delta_{n'l'}}{2\sqrt{(e_{n'l'}-\mu)^2 + \Delta^2_{n'l'}}},
  \label{D-nl}
\end{equation}
where the sum over $m$ has been performed because we  consider a spherical box where the gap does not depend on the azimuthal quantum number. In the limit where one can consider only diagonal matrix elements, the gap equation boils down to

\begin{equation}
  \Delta_{nl} = -g(l+1/2) v_{nl,nl}= -g(l+1/2)\int d^3r\rho_{nl}^2(r),
\end{equation}
where $g$ is the coupling constant and  $\rho_{nl}(r)$ is the density which corresponds to the s.p level with quantum numbers $nl$. It is clear that the density for the box potential is expressed by the square of spherical Bessel functions.
We have repeated the calculation of Croitoru {\it et al.} \cite{Shanenko11} for this gap as a function of grain radius taking the same values of the parameters for Sn.
In Fig.\ref{gapzoome} we show the averaged gap $\Delta=\sum_{nl}\Delta_{nl}u_{nl}v_{nl}/\sum_{nl}u_{nl}v_{nl}$  as a function of the grain radius $R$ in the range between 3.0 nm and 3.2 nm ($u_{nl}v_{nl}=\Delta_{nl}/[2\sqrt{(e_{nl}-\mu)^2 + \Delta_{nl}^2}]$ is the pairing tensor strongly peaked around the chemical potential). We clearly see how one level after the other enters and leaves the Debye window. This is indicated by the well delimitated and almost horizontal plateau's.

We, thus, see that a diagonal approximation is well justified for small sizes of a couple of nm.

Our aim is now to find a semiclassical approximation for the matrix element at the Fermi level. 
To this end, we can express the argument of the spherical Bessel functions by the Local Density Approximation (LDA). That is

\begin{equation}
  \rho_{nl}(r) \rightarrow \rho_{\mu l}(r) = \frac{1}{g_{\mu l}}j_l^2(k_F(r)r),
  \label{jl-lda}
\end{equation}
where $g_{\mu l}$ is the corresponding DOS obtained by integrating $\rho_{\mu l}(r)$ over $r$. The local Fermi momentum is given by $k_F(r) = \sqrt{\frac{2m}{\hbar^2}\left(\mu -V(r)\right)}$ where $V(r)$ is the mean field potential. Then for the box we have $k_F(r) = k_F\Theta(R - r)$ where $R$ is the radius of the box. Thus the local density on the Fermi level boils down to the {\it Fermi gas approximation} (FGA)

  \begin{equation}
\rho_{\mu l}(r) = \frac{1}{g_{\mu l}}j_l^2(k_Fr)\Theta(R-r).
  \end{equation}
  One can check that this FGA averages, as a function of $R$, very well the fully quantal matrix elements. This FGA has, however, the drawback that the integrals of Bessel functions for high values of $l$ are not completely easy to handle and, additionally, the $l$ values are still quantized. So one can suppose that there exists a more basic semiclassical approximation which will be a continuous function of $l$.


    
\begin{figure}
  \begin{center}
    \includegraphics[width=7.5cm]{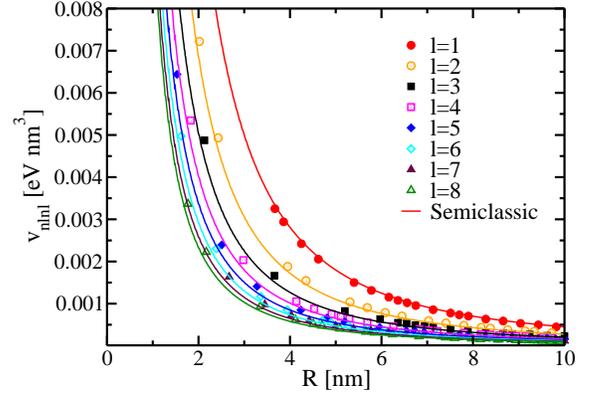}
    \caption{Semiclassic (solid line) and quantal matrix elements (symbols) as a function of $R_c$ for different values of $l$ with single parameter $d$ in expression (\ref{me-l}).}
    \label{me-fit}
    \end{center}
  \end{figure}

Indeed, the FGA can be reduced to such a semiclassical expression for the density of individual levels (the 'on the energy 
shell density') as was derived by Hasse in \cite{Hasse}

\begin{eqnarray}
  \rho_{E,l}( r) &=& \frac{l}{8\pi^2r^2\tilde g_{E,l}}\bigg ( \frac{2m}{\hbar^2}\bigg )^{1/2}\bigg [ E-V(r) - \frac{\hbar^2l^2}{2mr^2}\bigg ]^{-1/2}\nonumber\\
  &\times&\Theta\bigg ( E-V(r) - \frac{\hbar^2l^2}{2mr^2} \bigg ),
\end{eqnarray}
where $\tilde g_{E,l}$ is the corresponding level density obtained from above expression for the on the energy shell density by 
integration $\tilde g_{E,l} = \int d^3r \rho_{E,l}({\bf r})$, that is

\begin{equation}
  \tilde g_{E,l} = \frac{l}{2\pi E}\sqrt{k_E^2R^2 - l^2},
\end{equation}
where $k_E = \sqrt{2mE/\hbar^2}$ is the momentum at energy $E$.\\
\noindent
With this the semiclassical matrix element writes

\begin{equation}
  v_{El,El} = g\int d^3r\rho_{E,l}^2(r).
  \label{v-El}
\end{equation}
Unfortunately, this integral diverges at the lower limit. This often happens with semiclassical expressions which have to be considered as distributions when they are evaluated to some power (here power 2) and one has to regularize. 
We do this simply in calculating this matrix element in introducing a lower cut off $d$. This yields

\begin{equation}
  v_{\mu l,\mu l} = g\frac{1}{8\pi}\frac{k_F^3}{k_F^2R^2-l^2}\frac{1}{l}\ln \left[\frac{(Rk_F-l)l}{k_Fd(Rk_F + l)}\right].
    \label{me-l}
\end{equation}
For $d=0.003$ nm, we obtain very good agreement with the quantal values for various values of $l$ with this single parameter $d$, 
 as it can be seen in Fig.\ref{me-fit}. The agreement deteriorates slightly for higher $l$-values, but at the same time the values of the matrix elements decrease so that the slight inaccuracy is not very important.
Experimentally, one cannot measure the gap for each $l$-value separately. Only some average at the Fermi energy is accessible. For the quantal gaps, we do that in taking an average with the already used pairing tensor $\kappa = uv$ which is peaked at the Fermi surface. Semiclassically we replace $l$ by its mean value $\langle l\rangle$ so we get

\begin{equation}
    \langle l \rangle = \frac{ \int_0^{k_FR} l dl  }{k_FR}  = \frac{1}{2}k_FR .
    \end{equation}
This gives the following expression for the matrix element
\begin{equation}
  v_{\mu l,\mu l} = \frac{g}{3\pi}\frac{1}{R^3}\ln\left(\frac{R}{6d}\right).
  \label{v-R}
\end{equation}


\noindent   With this, we now calculate the semiclassical gap in diagonal approximation as

  \begin{equation}
    \Delta = \frac{\langle l\rangle}{3\pi}\lambda \frac{1}{N(0)}\frac{1}{R^3}\ln\left(\frac{R}{6d}\right),
    \label{diag-gap}
  \end{equation}
  where $\lambda = gN(0)$ and $N(0)=\frac{1}{4\pi^2}\frac{2m}{\hbar^2}k_F$ is the usual expression for the DOS at Fermi.
  In our numerical example we take the parameters for Sn \cite{Shanenko11}: 
$gN(0)=0.25;~~k_F = 16.4$ nm$^{-1}$; $\hbar^2/(2m) = 38.1$ meV nm$^2$, leading to a bulk value of the pairing gap of $\Delta_B=0.61$ meV.

\begin{figure}
  \includegraphics[width=7.5cm]{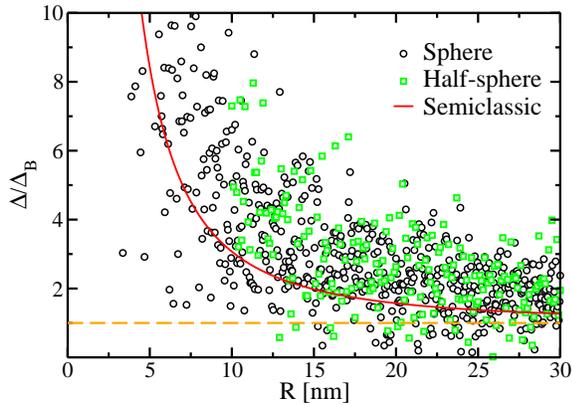}
  \caption{Gap for the sphere as a function of radius $R$ (circles) is shown together with the semiclassical result. The squares correspond 
to the hemisphere with the same radius $R$. Why sphere and semi-sphere give very similar results is explained in the text. The broken yellow line indicates the asymptotic value.}
    \label{gapZOOM}
\end{figure}

  Since at large radii the diagonal approximation goes to zero, we simply added the bulk value $\Delta_B$ of the gap to the expression (\ref{diag-gap}) as

  \begin{equation}
    \frac{\Delta}{\Delta_B} = \frac{\langle l\rangle}{3\pi}\frac{gN(0)}{\Delta_B}\frac{1}{N(0)}\frac{1}{R^3}\ln\left(\frac{R}{6d}\right) + 1.
    \label{D/Db}
    \end{equation}
For very small values of the radii, this constant addition is of little importance, since there the gap diverges.
There will be some error in the transition region where the diagonal approximation goes over smoothly into the bulk value. We estimate 
that the local error may be around 10-15 \%. It would eventually be possible to model the transition region more accurately, however 
with a much heavier formalism. Since, for the average behaviour, we are more interested in a qualitative, semi-quantitative description, 
we refrained from this additional effort.
In Fig.\ref{gapZOOM}, we see quite reasonable agreement between the quantal result for the sphere and the semiclassical approximation (\ref{D/Db}).

\begin{figure}
  \includegraphics[width=7cm]{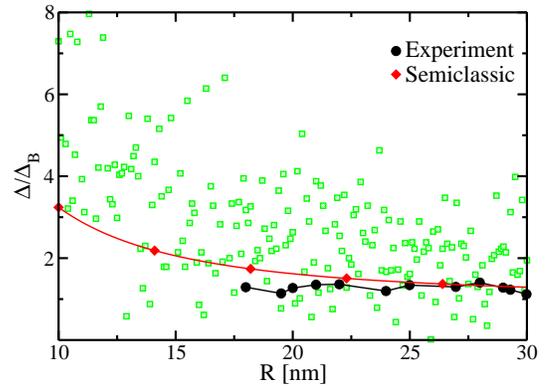}
  \caption{Gap in a size-range (empty squares) where some experimental values exist (full circles). The straight lines connecting the points are to guide the eye.}
    \label{Garcia}
\end{figure}

In Fig.\ref{gapZOOM}, we also show the values for the hemisphere. Actually sphere (circles) and hemisphere (squares) give very similar results. This stems from the fact that the  level density is half of the one of the sphere. However, also the volume is half, so that both effects cancel. A detailed study of the hemisphere is given in  Appendix \ref{appendix3}.
For the hemisphere exist some experimental values in Ref.\cite{G-G} in the range 18 nm $<$ $R$ $<$ 30 nm. Actually it is not explained in \cite{G-G} from where the experimental values, which are supposed to be average values, are obtained. In the original paper by Bose {\it et.al.} \cite{Nature}  the values are always shown with respect to the average, however, the average is never displayed. 
In Fig.\ref{Garcia}, we show that there is reasonable agreement between the quantal  and semiclassical results together with the experimental values given in \cite{G-G}.

\begin{figure}
  \includegraphics[width=7.cm]{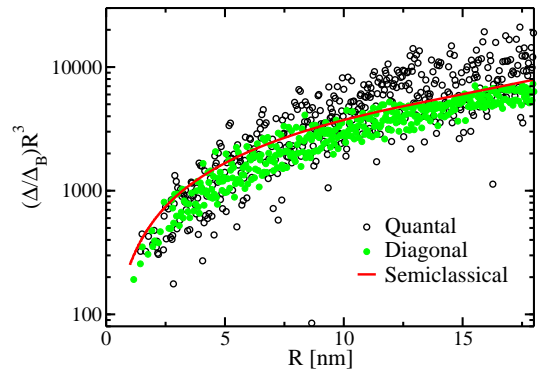}
  \caption{\label{log}The size dependence of the average gap where the  volume dependence is divided out.
    The fluctuating  curve is the quantal result and the smooth curve the semiclassical result.}
\end{figure}

One may wonder whether the logarithmic function in the semiclassical matrix element is real or an artefact of the approximation.
In Fig.\ref{log} we, therefore, show the quantal gap of metallic grains (empty circles) shown already in Fig.\ref{gapZOOM} with the diagonal approximation of the matrix element alone (full circles) together with semiclassical curve (solid line), however all multiplied with $R^3$ in order to divide out the volume dependence and slightly normalised to the value at the origin of the figure. This allows us to appreciate the influence of the $\ln(R/6d)$ term alone. We see that there is a genuine logarithmic dependence in the average quantal matrix elements.

Let us conclude concerning the results of electronic nano-devices. Besides for films, the semiclassical Weyl description is faced with some difficulties stemming from the fact that for system-sizes of only a few nanometers the Debye window lets penetrate, as a function of varying sizes, discrete levels only one by one, rendering the system highly quantal. For small metallic grains this difficulty could be circumvented because what counts in such cases is finally the size dependence of the pairing matrix elements for which good semiclassical expressions exist. The case of metallic nano-wires is intermediate. Discrete transverse levels are embedded in the continuum of the infinite wire. This makes a semiclassical approach particularly demanding because the quantal fluctuations around the average go mostly to one side and not symmetrically to both sides as in the case of films. We could find a very accurate global correction factor to the semiclassical expression with which the quantal results are perfectly averaged. This then also yields very good reproduction of the existing experimental data. 
In the case of films the quantal fluctuations are symmetric around the average and small enough, so that the semiclassical Weyl approach yields a perfect average of the fluctuating gaps. Then the increase of the gaps with decreasing sizes is clearly due to the increase of the influence of the surface pressure giving rise to an increase of the DOS. We will now consider the situation of pairing of another ultra-small Fermi-system: atomic nuclei.

\section{Applications of the semiclassical BCS approach to nuclear systems}\label{sec:nuclei}
\subsection{Generic applications}
\subsubsection{The box potential case}

As we mentioned in the introduction, the BCS approximation works best for potentials with a steep surface.
Therefore, let us first consider, for a schematic study, the extreme case of a spherical box with infinite walls also for nuclei. With respect to the metallic devices, in the nuclear case, the main difference is that there is no Debye window very narrowly concentrated around the Fermi level. On the other hand, in nuclear systems, the pairing force is, in principle, finite range \cite{Gogny}. However, also in nuclear physics {\it effective} zero range pairing forces are in use \cite{bertsch91,garrido99}. In the latter case the divergency of the gap is either handled with a cut-off or by regularization \cite{Bulgac}. We  employ here the latter possibility. We follow the procedure of Grasso-Urban \cite{Grasso}; the HFB equations for a finite system with zero range force are then given by

\begin{equation}
  \Delta_{HFB}(r) = \frac{1}{2}v_{eff}(r)\sum_{\alpha}U_{\alpha}(r)V_{\alpha}(r),
  \label{nucl-hfb}
\end{equation}
where the amplitudes $U, V$ are the standard ones of the HFB or BdG equations. The effective position dependent coupling is given by

\begin{equation}
  v_{eff}(r) = \bigg [ \frac{1}{v_0}- \frac{mk_C(r)}{2\pi^2\hbar^2}\bigg ( 1- \frac{k_F(r)}{2k_C}\log\frac{k_C + k_F}{k_C-k_F} \bigg) 
\bigg ]^{-1},
  \label{veff}
\end{equation}
where $v_0$ is the bare coupling constant and $k_C$ is a cut-off parameter on which the result should not depend in a certain range around the Fermi energy.
For the BCS case, we use a modified version

\begin{equation}
  \Delta_{BCS}(r) = \frac{1}{2}v_0\sum_{\alpha}\bigg [U_{\alpha}(r)V_{\alpha}(r) - \frac{\Delta_{\alpha}\phi_{\alpha}^2(r)}{2\varepsilon_{\alpha}} \bigg ],
    \label{nucl-bcs}
\end{equation}
where $\phi_{\alpha}(r)$ is the single particle mean field wave function and $\varepsilon_{\alpha}$ is the single particle energy 
corresponding to the (self-consistent) mean-field potential. The parameter $v_0$ in (\ref{nucl-bcs}) has been adjusted in such a way that in infinite matter it is equal to $4\pi \frac{\hbar^2}{m}a$, see Eq.(14). Therefore, in infinite matter both expressions (\ref{nucl-bcs}) and (\ref{veff}) give the same result, that is $v_0$ has been adjusted in this way. Its value turns out to be: $v_0$ = 128.28 MeV fm$^3$. By extension, we keep (\ref{nucl-bcs}) also for finite systems.\\
In the case of the box the lowest order TF-BCS approximation gives a constant gap from the center to the border of the box.
This gap depends on the Fermi energy, or equivalently on the chemical potential and corresponds to the value in the infinite 
system, i.e. to its bulk value. It is the Fermi-gas approximation to the gap for finite nuclei.

To obtain the gap as a function of the distance inside the box, we have to fold the energy dependent gap with the radial dependent spectral 
density. We  explain how this can be achieved.
The starting point is the on-shell distribution function, i.e. the Wigner transform $f({\bf R},{\bf p})$ of the single-particle density matrix 
averaged over the energy shell, which has been given in section II, Eq.(\ref{Weyl}).
The local part of the pairing tensor in coordinate space is thus obtained integrating $\kappa(E)$ weighted with the local level 
density $g(E,{\bf R})=\int \frac{d^3p}{(2\pi \hbar)^3}f({\bf R},{\bf p})$ derived in Appendix \ref{appendix1} as
\begin{eqnarray}
\kappa({\bf R}) &=& \int_0^{\infty}dE g(E,{\bf R}) \kappa(E) \nonumber\\
&  =&\int_0^{\infty} dE g_0(E) \kappa(E) \big(1 - j_0(2R p_E)\big),
\label{kappaR}
\end{eqnarray}
where we have used $p_{E} = \sqrt{\frac{2mE}{\hbar^2}}$ and $g_0(E)=\frac{1}{4 \pi^2}(\frac{2m}{\hbar^2})^{3/2}\sqrt{E}$.
Eq.(\ref{kappaR}) fulfils $\kappa(R=0) = 0$ and $\kappa(R=\infty) = \kappa_0$, where $\kappa_0$ is the pairing tensor in the 
infinite system. The box surface is taken as the origin. The gap as a function of $R$ is then simply given by $\Delta(R) = - g \kappa(R)$. 
 To account for finite size, i.e., quantal effects, we multiply this result by the ratio 
 between the gap at the Fermi energy and the gap in the bulk $\Delta(R)/\Delta_B$. This ratio is derived in Appendix \ref{appendix2} 
and is given by

\begin{equation}
\frac{\Delta(\mu)}{\Delta_B} = \bigg (1+\frac{\pi}{4k^B_F}\bigg )e^{\frac{4\pi^2\hbar^2}
{v_02mk^B_F}\frac{\pi}{8k^B_F}\frac{S_M}{V_M}}e^{\frac{\pi}{2k^B_F}
    \frac{S_M}{V_M}(\ln 2 - 1)},
  \label{finite-size}
\end{equation}
where $V_M$ and $S_M$ are, as in the case of metallic nano-systems, matter volume and matter surface, which correspond to an
effective box radius enclosing the right number of particles in the Fermi-gas limit.

\begin{figure}
\includegraphics[width=6.5cm,angle=0]{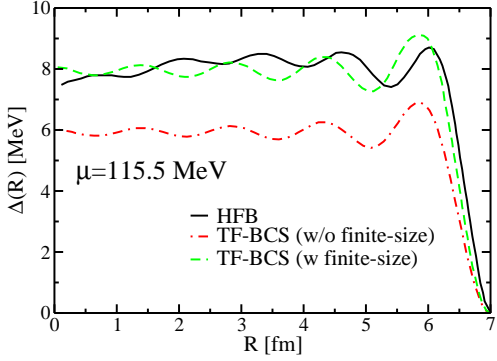}
\caption{\label{gap-box} Local gap with zero range force and regularisation in spherical box potential with HFB (solid line)  
and TF-BCS with finite-size correction (dashed line) and without (dashed-dotted line).}
\end{figure}

We see in Fig.\ref{gap-box} that there is again good agreement between the TF-BCS (dashed line) (multiplied with the enhancement factor 
$\Delta(\mu)/\Delta_B$, see Eq.(\ref{finite-size}) ) and quantal HFB (solid line) results.
Notice that for the sake of clarity, we have reversed the horizontal scale in this figure putting the radius of the box at the surface.
 The semiclassical TF-BCS calculation, obtained starting from Eq.(\ref{Weyl}), reproduce reasonably well the quantal oscillations 
close to the surface.
In the interior the 
oscillations get out of phase. This is not astonishing, since we only gave information around the surface as input. 
Such behavior is known from other situations \cite{me-paper}.  In  Fig.\ref{gap-box} the dashed-dotted line 
corresponds to the TF-BCS gap without the enhancement factor due to the finite-size.
For the force we used zero range with the strength constant $v_0$ = -128.28 MeVfm$^3$.\\

\begin{figure}
\includegraphics[width=8cm]{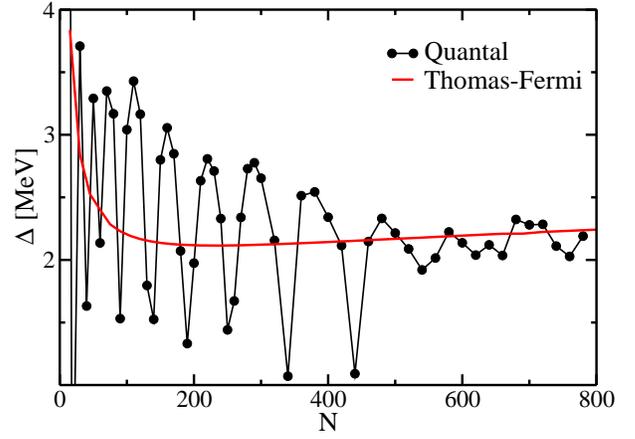}
\caption{\label{gap-box-N} Gap with zero range force and regularisation in spherical box potential with a fictitious radius of $R=A^{1/3}$ fm calculated with HFB (black full dots), and TF-BCS (continuous line, red) along the stability line as a function of N. See text for details.}
\end{figure}

We also calculated in a large box, and as an academic example, the nuclear gap  at the Fermi energy as a function of the neutron 
number $N$ along the line of stability of nuclei with box radius $R = A^{1/3}$, which is displayed in  
Fig.\ref{gap-box-N} computed with the HFB and TF-BCS approximations. 
In the semiclassical case,  we use the semiclassical formula (\ref{finite-size}) with $S_M/V_M = 2/R$. The relation between the neutron number
$N$ and mass number $A$ is provided  by the stability condition along isobaric parabolae \cite{stability}

\[A-N=A/(1.98 + 0.0155A^{2/3}),\]
We again see good agreement between HFB and the TF-BCS results.



The semiclassical gap can also be obtained at finite temperature in TF-BCS approximation. Equation (9) is then modified in the following 
way

\begin{equation}
  \Delta_T(E\!) =\!\! \int_0^{\infty}\!\!dE'g^{\rm TF}(E')V(E,E')\kappa(E')(1-2f(E_{\rm q.p.}(E')),
  \end{equation}

\begin{figure}
  \includegraphics[width=8cm]{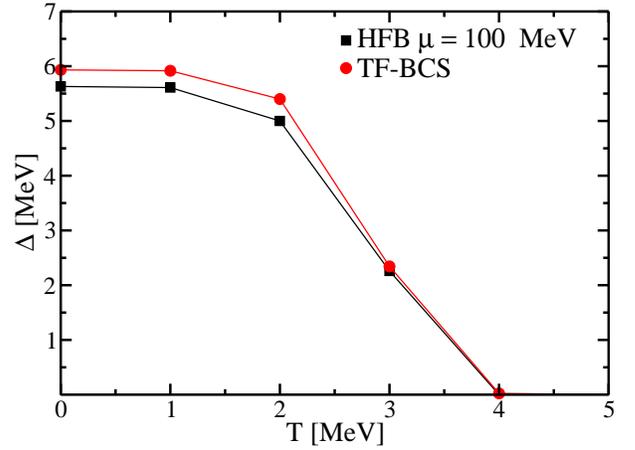}
  \caption{\label{gap-T} Gap in box with $R_c$=7fm with $v_0 = -128.28$ MeV as a \
    function of the temperature
    computed with HFB (squares) and TF-BCS (circles)}
\end{figure}

The gap in the box with radius 7 fm is shown in Fig.\ref{gap-T} as a function of temperature for $\mu$ = 100 MeV. We have chosen this high chemical potential because it is at the limit where shell effects disappear. Again we see very close agreement between HFB and BCS results. That even the critical temperature is accurately obtained with BCS highlights again the fact that BCS is, for steep mean-field potentials, a very valid approximation. For example in the case of a harmonic potential, this is not at all the case, see \cite{Grasso} and the BCS approximation fails completely there (see discussion concerning Fig.1 in Sect. II).

\subsection{Woods-Saxon or Fermi-function types of potentials}

After these first examples concerning again the box potential, let us consider a much more realistic nuclear situation. First of all, 
at least in principle, the nuclear pairing force is finite range. A very well studied and successful effective force of this type is the 
so-called Gogny force \cite{Gogny}. We  will consider here the D1S version. With respect to zero range forces, finite range interactions
  complicate enormously the solution of the pairing problem, that is the solution of the HFB equations. Because of 
this complication, in nuclear physics also effective zero range or separable forces \cite{tian09} are in use, see, e.g., Ref.\cite{garrido99} 
where a zero range force is derived from the Gogny force which has become quite in use in the nuclear physics community. However, the most realistic calculations are done with the 
mentioned Gogny force which we will use in our study below. Also in self-consistent HFB calculations the corresponding mean-field potential is 
finite, similar to a Fermi function. With respect to the box potential, this will introduce quite significant changes as we will see in a 
moment. Not to complicate the study too much, we will not consider the self-consistent Gogny mean-field potential, but replace it by a 
phenomenological one. This should be fully sufficient for the effects we want to study.
A typical phenomenological Woods-Saxon potential of nuclear physics
which is often used is given by the so-called Shlomo potential \cite{shlomo91} 
\[ U(r) = \frac{U_0}{1+e^{(r-R_0)/a}},\]
  where the parameters are defined as
\begin{eqnarray*}
   U_0 &=& -54 + 33 \frac{N-Z}{A},\\
   a&=&0.70\; {\rm  fm},\\
   R_0 &=&R_V/[1+(\pi a/R_0)^2]^{1/3},\\
   R_V &=& 1.12A^{1/3} + 1.0\; {\rm  fm} .
  \end{eqnarray*}

For our study, we choose an artificially large nucleus with $N=Z=1000$, which is a sufficiently big number
so that the arches (see below) corresponding to the shell effects are washed out. As a consequence of this large nucleus as a study object the chemical potential turns out to be $\mu= -16$ MeV. For nuclei, this is an academic example which, however, will allow us to discuss in a transparent way all relevant features induced by a finite mean-field with a finite range pairing force. We will come back to this discussion at the end of this section. For small superconducting metallic grains the example just presented may even be relevant (with parameters scaled to electronic devices) since, in principle, also there the mean field is Fermi function like \cite{Kohn}.\\
As a last point, we should mention that the mean-field with a finite range force is non-local implying the existence of an effective mass $m^*$. In nuclear physics an average value of the effective mass is $m^*/m \simeq $ 0.7. The Shlomo potential does not contain an effective mass, so we took the strength of the Gogny force scaled down to 85 percent. This simulates reasonably well the influence of the effective mass.

\begin{figure}
\includegraphics[width=7cm]{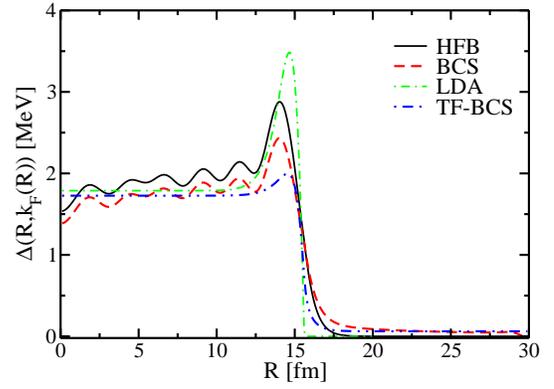}
\caption{Comparison between HFB (black) results for the radial dependence of the gap in a large system with 
$N=1000$ fermions in the Shlomo potential and the BCS (red) ones. The LDA (green) estimate and the semiclassical TF-BCS (blue) 
prediction are also shown.}
\label{Ale-Wigner-gap}
\end{figure}

\begin{figure}
\includegraphics[width=7cm]{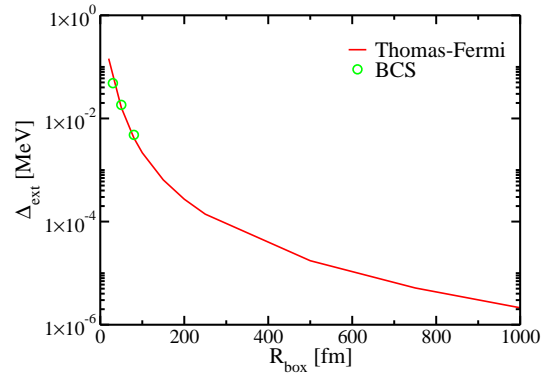}
\caption{The gap in the outer gas region as a function of the box size $R$. The TF-BCS solution is represented by the continuous line (red) 
while the quantal BCS  results are given by the circles (green). Notice the very fast disappearance of the gap with $R$!}
\label{gasTFBCS}
\end{figure}

In Fig.\ref{Ale-Wigner-gap}, we show $\Delta(R)$ with HFB, BCS,  TF-BCS, and LDA. 
We have to explain, how this local gap function is obtained. In general the gap is a non-local matrix in $r$-space, 
$\Delta({\bf r},{\bf r}')$, because the Gogny force is of finite range. This matrix is written in center of mass and relative coordinates 
and Fourier transformed with respect to the relative coordinate. This Wigner-transform of the gap matrix then yields a function in phase space:
 $\Delta({\bf R}, {\bf p})$. Quantally, this quantity depends on the angle of momentum ${\bf p}$ but here  in Fig.\ref{Ale-Wigner-gap}, 
it is avaraged over the angle and the replacement $p \rightarrow p(R)=\hbar k_F(R)$ is made with $k_F(R) =  \sqrt{\frac{2m}{\hbar^2}(\mu - U(R))}$,
 the local Fermi-momentum. This definition has the advantage that for very large systems the gap becomes the one of infinite matter taken at the Fermi surface. In the TF-BCS approach, in order to treat the same quantity, we consider the following expression

\begin{equation}
  \Delta(R,p) = \int \frac{d^3p'}{(2\pi \hbar)^3}v(p,p')\kappa(R,p'),
\end{equation}
with $p$ replaced by $p(R)$ and,  as discussed in section \ref{sec:TF},

\begin{equation}
  \kappa(R,p) = \int dE\kappa(E)\delta\left(E-\frac{p^2}{2m} - U(R)\right).
  \label{kappa-Wigner}
\end{equation}
However, since the local Fermi momentum terminates at the classical turning point, it is preferable to perform an average of 
$\Delta(R,p)$ over the Wigner transform of the anomalous pairing tensor $\kappa(R,p)$, which is in semiclassical approximation strongly 
peaked around $p(R)$. We, thus, consider for the local gap in TF-BCS approximation

\begin{equation}
  \Delta(R) = \int \frac{d^3p}{(2\pi \hbar)^3}\Delta(R,p)\kappa(R,p)/\kappa(R),
\end{equation}
where $\kappa(R) = \int \frac{d^3p}{(2\pi \hbar)^2}\kappa(R,p)$. Above expression can then be rewritten in the following form
\begin{widetext}
\begin{equation}
  \Delta(R) = \frac{1}{\kappa (R)}\bigg (\frac{1}{4\pi^2}\frac{2m}{\hbar^2} \bigg )^2\int dEk_E(R)\kappa(E)
\int dE'k_{E'}(R)\kappa(E')v(E,E';R);
\end{equation}
\end{widetext}
where 
\begin{eqnarray*}
2\kappa(E) &=& \Delta(E)/\sqrt{(E-\mu)^2 + \Delta^2(E)},\\
 p(R)=\hbar k_E(R) &=& \sqrt{2m(E - U(R))}, \\
v(E,E';R) &=& v(p=\hbar k_E(R), p'=\hbar k_{E'}(R)).
\end{eqnarray*}
This is the expression which we used in our numerical application.
In Fig.\ref{Ale-Wigner-gap} we show four lines corresponding to HFB (solid), BCS (dashed), TF-BCS (dashed-double dotted), and LDA (dashed-dotted). 
The first thing we notice, is that quantal BCS is globally about 10 percent lower than the full HFB solution. This may be a 
little more than what we have seen using the box potential but BCS can still be considered as a good approximation. In this 
respect we should remember that BCS is, numerically, a tremendous simplification over HFB. However, a close look into the outer region seems to reveal 
that there BCS develops a small but finite gap which is absent with HFB. Before we discuss this in detail, let us present the TF-BCS solution. 
It can also be considered as a valuable approximation though the surface peak is a little small and pushed more to the outside. However, 
the fall off in the surface region is quite reasonable joining very accurately the finite gap value in the outside region. This feature 
of BCS is actually well known in the nuclear physics community and considered as one of the major draw backs of BCS versus HFB. However, 
here we made a very careful study how the gas behaves as a function of the radius of the outer confining box, which in the 
Fig.\ref{Ale-Wigner-gap} is located at $R=30$ nm. Since quantally such large box sizes are difficult to manage, we first made a study 
with TF-BCS, which seems to grasp the situation in the outer region very well. In Fig.\ref{gasTFBCS} we show how the TF-BCS gap varies in the outer 
region as a function of the box radius. In the semiclassical approach it is no problem to go to very large box sizes of the order of 
thousand fm. To our surprise, we found that the gas disappears as a function of the box radius very quickly. At the same time we noticed that the gap 
inside is only very little affected (it is marginally reduced). We then repeated the study with quantal BCS. The largest box radius we 
could handle was $R=80$ fm. Though at this radius the gas is not yet zero, we found, however, that it was reduced in the same proportion 
as the TF-BCS gap, see 
circles in Fig.\ref{gasTFBCS}. The conclusion therefore is that the BCS solution also goes to zero in a similar fashion as HFB under the proviso that 
one takes a sufficiently large box size. We also found that the integral over the gas portion of the gap stays constant as a function of the radius of 
the limiting box. One may, therefore conjecture that  the particle number which is spilled out into the continuum remains a constant 
independent of the size of the total system. This is a new insight into the BCS solution, despite the fact that in practice large box 
sizes may not be easy to handle. But given the already mentioned tremendous simplification of BCS over HFB, this is an interesting 
aspect.  Let us shortly come back to the relevance of our model with respect to realistic situations. We also considered our model in 
the asymmetric case with much more neutrons ($\mu_n = -3.4$ MeV; N = 1250) and $ Z$ = 750. Since now the chemical potential is much closer to the upper edge of the potential, one may suppose that the spill out into the continuum is much more important. The gap in the gas phase becomes indeed larger but in relatively modest proportions. The effect that the gas disappears with increasing box size persists though the convergence is slightly slower. It seems clear that this is a generic effect. We, thus, can conclude that our model represents with respect to the difference between BCS and HFB all the characteristic features which are encountered in realistic situations.


We also show in Fig.\ref{Ale-Wigner-gap}  the corresponding LDA solution which overshoots very much the surface peak and stops at the classical turning point. 
Actually this oversized surface peak is here still relatively moderate. There can be situations where peak hight is still more pronounced.

In conclusion of this section, we can say that even in realistic cases with finite range pairing forces and Woods-Saxon type of 
potentials, the gap in BCS approximation is, with respect to HFB, reduced but only of the order 10-15 percent. In view of the 
very strong simplification of BCS with respect to HFB, this is a quite acceptable approximation. This the more so as one can remedy 
for this slight underestimation by increasing the intensity of the pairing force a little. This point of view has, e.g., been 
adopted by Ring {\it et al.} \cite{RingBCS}. These authors conclude that a 6 percent increase of the intensity of the pairing force is sufficient to recover from BCS the HFB situation. As mentioned already and as a word of caution, let us say again that this favourable situation 
for BCS may nevertheless deteriorate for exotic nuclei when coming very close to the drip-line where the last neutrons are almost unbound. This scenario may happen for nuclei in the crust of neutron stars and in nuclei created in supernovae.

\subsection{Recovering shell effects}

\begin{figure}
\includegraphics[width=7.cm,angle=0]{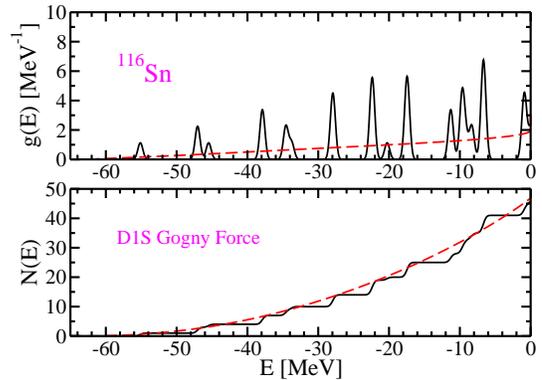}
\caption {\label{av-dos}Slightly smoothed level density and corresponding number of states obtained using a quantal calculation (solid line) and TF-BCS (dashed). See text for details.}
\end{figure}

\begin{figure}
\includegraphics[width=7.5cm,angle=0]{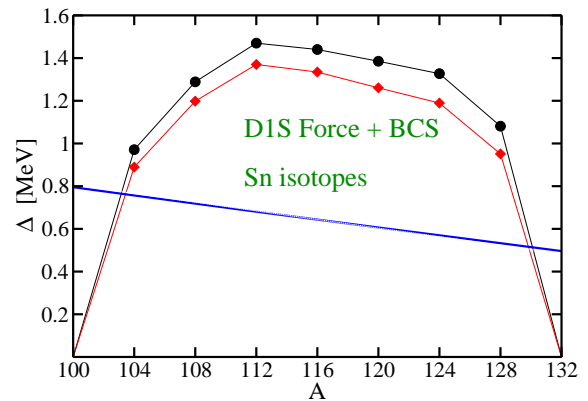}
\caption {\label{recov-quantal}Solution of TF-gap equation  with quantal level density (full diamonds) as compared to the quantal BCS results (full circles). The straight ( blue) line is the TF-BCS gap without recovering shell effects. It is seen that it averages the quantal gap.}
\end{figure}

As we know, the value of the gap is very much dependent on the level density. Most of the shell effects are contained in this level density. 
The shell fluctuations of the pairing matrix element are less important \cite{sc-me}. It should, therefore, be a reasonable approximation to insert the 
quantal level density into the TF gap equation (\ref{D(E)},\ref{kappa(E)}). 
To this end we start from the quasi-local reduction of the HF energy density associated to a finite-range
interaction \cite{soubbotin00,soubbotin03}, which allows to write the quantal  mean-field and effective mass in a local form.
Once the single-particle levels of this quantal mean-field have been obtained, we build the fluctuating level density by folding the 
quantal level density with a Gaussian with a width $\sigma$=0.5 MeV and with a strength such that the area below the Gaussian equals
the degeneracy of each level (spherical symmetry is assumed). This method is known in nuclear physics as the Strutinsky averaging \cite{RS}.
The smooth level density and the accumulated number of states obtained in this way for the nucleus $^{116}$Sn using the D1S force 
are displayed in the upper and lower panels, respectively, of Fig.\ref{av-dos}. The average quantal gaps \cite{krewald06} weighthed
with $u^2v^2$ (circles) and the ones obtained with the TF-BCS theory using the fluctuating level density (diamonds) are displayed in
Fig.\ref{recov-quantal}. We see that by introducing the smoothened level density the quantal arch structure is recovered and
 the semiclassical gaps obtained in this way reproduce rather well the quantal values. In the same figure, we also display the 
TF-BCS averages of the gap computed with TF level density (full straight line). In this case the quantal arch structure is washed out and the
semiclassical average gaps show a downward trend with increasing mass number (see also \cite{IJMP2011,50yearsBCS} for more details).

This may be a promising and good approximation for situations where even the quantal solution of BCS becomes computer time consuming as, e.g., for heavy (tri) axially deformed nuclei.

\subsubsection{ Average gaps as a function of mass number and asymmetry in comparison with realistic calculations}

\begin{figure}
  \includegraphics[width=7.5cm]{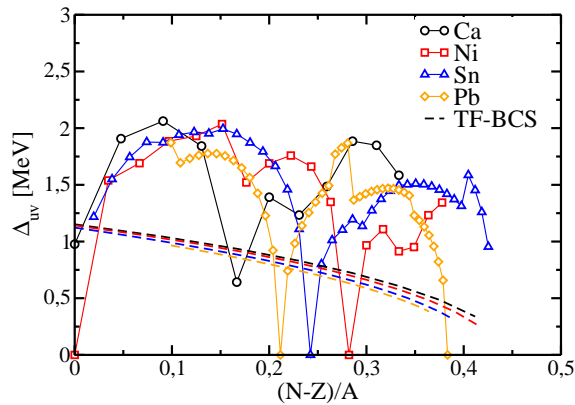}
  \caption{BCS-Gaps as a function of the asymmetry for Ca, Ni, Sn and Pb isotopic chain with the Shlomo MF-potential 
and the re-scaled Gogny D1S pairing force. The broken lines are the corresponding TF-BCS curves.} 
  \label{gapsI}
\end{figure}

Since we have seen that BCS is also a valid approximation for realistic mean-field potentials, we go on and show 
in Fig.\ref{gapsI} with the Shlomo potential and the D1S Gogny force  in the pairing channel the evolution of the gap values as a function 
of mass number $A$ and asymmetry $(N-Z)/A$. 
The gaps are calculated with quantal BCS and also with TF-BCS. The D1S force has been downscaled as before to 85 percent, since the 
Shlomo potential does not contain an effective mass which reduces the gap \cite{50yearsBCS}. The quantal calculations are sufficiently 
realistic so that they stand  well for the experimental values. They are not shown not to overload the figure but can be seen in Fig.1 of 
\cite{pastore13}.  We see, on the one hand, that the TF-BCS results represent well 
the average of the quantal ones, and, on the other hand, that there is a downward trend with increasing $N$ in both approaches. 
Actually  we have shown that the gap is very much reduced when, e.g.,  approaching the neutron drip and even beyond when the continuum is populated with a neutron gas as is the case in neutron stars, see \cite{pastore13}.\\

We also want to mention that the TF-BCS gaps are very well approximated with a pocket formula (PF) analogous to (\ref{finite-size}) but with a finite range (Gogny) force for which no regularisation is necessary. It is given by

  \begin{equation}
    \Delta(\mu) = 8\mu e^{-\frac{1}{g(\mu)V(\mu,\mu)}}F(U_0/\mu)
    \label{PF-F1}
  \end{equation}
  with

  \[ F(U_0/\mu)=e^{\sqrt{|U_0|/\mu}-2} \sqrt{\frac{ \sqrt{|U_0|/\mu}-1}{\sqrt{|U_0|/\mu}+1}}.\]
    The chemical potential should here be counted from the bottom of the mean field potential. In weak coupling this then yields $\mu = \varepsilon_F$. This formula is interesting, since it immediately shows that the average gap breaks down towards the drip line.

    The derivation of (\ref{PF-F1}) is given in App.2

This ends the considerations of of the semiclassical approach to finite nuclei. 
We have demonstrated that also with finite potentials of Fermi-function shape the BCS approximation is quite reliable since the 
surface is steep. Then also the TF-BCS approximation represents the average behavior very accurately. This holds for relatively 
stable nuclei. Considering very exotic neutron rich nuclei, the BCS approximation may become less reliable. Many more details and 
results could be shown for nuclei which, however, would go beyond the scope of this paper. They will be  eventually given in a future specific nuclear paper. 

\section{Conclusions}\label{sec:conlcusion}

In this work we have seen that first of all, the BCS approximation is a quite valid approximation for 'steep' mean field potentials 
where the surface width is small compared to the total extension of the system as this is the case in most 'realistic' systems as, 
e.g., electronic nano-devices and nuclei. Further, we have demonstrated that semiclassical methods, as the Weyl approximation for box 
potentials and the Thomas-Fermi approximation for finite potentials with a distributed surface, are very useful tools to reveal generic average surface and size dependences of systems so largely different as are nano-sized electronic devices and atomic nuclei. Of course, in those ultrasmall systems strong shell effects are generally present. But eventually they can be added perturbatively and are, in any case, {\it not} responsible for the underlying, sometimes quite subtle, surface and size dependences. For instance one of the main agents of the increase of pairing with decreasing sizes is the fact that the increasing influence of the surface pressure with decreasing size compresses the systems and, thus, the Fermi momentum, i.e. the level density, increases and, consequently, the gap. This is very clearly born out in very thin electronic films where quantal shell effects are not very pronounced. As soon as the dimensions are further reduced as in electronic wires and grains where discrete levels are involved, the quantal fluctuations raise very strongly. In such cases a new aspect has to be considered, namely that the Debye window is usually so small that for very small diameters of the systems, it may happen that only one or at most very few discrete levels enter the Debye-widow at once. This naturally becomes then a highly quantal situation where semiclassical methods are not very well adopted. Fortunately in these cases the gap field becomes practically diagonal and the main agent of the size dependence of pairing is given solely by the size dependence of the pairing matrix elements \cite{Shanenko11}. In earlier works, we have shown that the average size behaviour of the pairing matrix elements can again be very well reproduced with Thomas-Fermi like approaches \cite{sc-me}. Indeed, we could show that the average  size dependences of gaps of spherical grains as well as of grains of hemispherical shapes can very well 
be reproduced with our semi-classical methods.\\
As another superfluid system of ultrasmall size, we also briefly considered the case of atomic nuclei. Despite of the fact that there the pairing force is of finite range complicating very much the solution of the pairing problem, we were able to show that even realistic  HFB calculations for nuclear systems could, firstly be replaced to good approximation by corresponding BCS ones and, secondly, the average  BCS results were very well reproduced by our Thomas-Fermi calculations.
Average behaviour of pairing in nuclear systems is very valuable because it sheds light on the underlying physical reasons for its behaviour which quantally is very difficult to detect because of very strong  shell effects present in nuclei. A more detailed account of the nuclear situation may be given elsewhere.

\section{Acknowledgments}
We thank M. Farine for early contributions to this work. M. Urban is greatly acknowledged for critical comments concerning the use of BCS versus HFB approaches.
The work of X.V. was partially supported by Grant FIS2014-54672-P from MINECO and FEDER,
Grant 2014SGR-401 from Generalitat de Catalunya,
and Project MDM-2014-0369 of ICCUB (Unidad de Excelencia Mar\'{\i}a de Maeztu) from MINECO. The work of A.P. was supported  by STFC Grant No. ST/P003885/1.
This project was undertaken on the Viking Cluster, which is a high performance compute facility provided by the University of York. We are grateful for computational support from the University of York High Performance Computing service, Viking and the Research Computing team.

\begin{appendix}

\section { Weyl spectral density, density of states and pairing matrix elements}
\label{appendix1}
In this Appendix, we derive the spectral density, the density of states and the pairing matrix elements,  
which are used in different parts of this work, starting from the Weyl expansion of the distribution function
in a spherical box given by Eq.(\ref{Weyl})
\begin{widetext}
\begin{equation}
f_{E}({\bf R},{\bf p}) =
\frac{1}{g(E)}\bigg[ \delta\big(E - \frac{\hbar^2 p^2}{2m}\big) -
  \frac{2m}{\hbar^2} \delta p_z \frac{\cos(2 R k_{E}(p_x,p_y))}{k_{E}(p_x,p_y)} \bigg]
\Theta(R_0(\theta,\varphi)- R(\theta,\varphi))
\label{eqA41}.
\end{equation}
\end{widetext}
By assuming cylindrical symmetry, we can write the momentum along the $z$-axis as a function of total
momentum related to the energy $E$ and the momentum in the perpendicular plane as 
\begin{equation}
p_z = k_{E}(p_x,p_y) = \sqrt{p_{E}^2 - p_{\perp}^2},
\end{equation}
from where
\begin{widetext}
\begin{eqnarray}
g(E,{\bf R}) &=& \int f_{E}({\bf R},{\bf p})\frac{d{\bf p}}{(2\pi)^3} =
\int \frac{d{\bf p}}{(2\pi)^3} \delta\big(E - \frac{\hbar^2 p^2}{2m}\big)
- \frac{2m}{\hbar^2} \int \frac{d{\bf p}}{(2\pi)^3}\delta p_z \frac{\cos(2 R k_{E}(p_x,p_y))}{k_{E}(p_x,p_y)} 
\nonumber \\
&=& \frac{4\pi}{8\pi^3}\int dp p^2 \frac{m \delta(p-p_E)}{\hbar^2 p_E}
- \frac{2m}{\hbar^2} \frac{2\pi}{8\pi^3}\int p_{\perp} dp_{\perp}dp_z \delta p_z 
\frac{cos(2 R {\sqrt{p_{E}^2 - p_{\perp}^2})}}{\sqrt{p_{E}^2 - p_{\perp}^2}}
\nonumber \\
&=& \frac{1}{4\pi^2}\frac{2m}{\hbar^2}\bigg(p_E - \int_0^{p_E} p_{\perp} dp_{\perp} 
\frac{\cos(2 R {\sqrt{p_{E}^2 - p_{\perp}^2})}}{\sqrt{p_{E}^2 - p_{\perp}^2}}\bigg),
\label{eqA42}
\end{eqnarray}
\end{widetext}
and performing the change of variable $k=\sqrt{p_{E}^2 - p_{\perp}^2}$ becomes
\begin{eqnarray}
g(E,{\bf R})&=&\frac{1}{4\pi^2}\frac{2m}{\hbar^2}\bigg(p_E - \int^0_{p_E} dk \cos(2kR)\bigg)\nonumber\\
&= &\frac{1}{4\pi^2}\frac{2m}{\hbar^2}p_E\left(1 - j_0(2R p_E)\right),
\nonumber \\
\label{eqA43}
\end{eqnarray}
where  $j_0(x)=\sin x/x$ is the zeroth-order spherical Bessel function.

Once the spectral density (\ref{eqA43}) has been obtained, the density of states can be easily
calculating by performing the integral over ${\bf R}$ assuming semi-infinite geometry 
\begin{eqnarray}
g(E) &=& \frac{1}{4\pi^2}\frac{2m}p_E\int \big(1 - j_0(2R p_E)\big)d{\bf R}\nonumber\\
&=& \frac{1}{4\pi^2}\frac{2m}{\hbar^2}p_E \bigg( V - S\int_0^{\infty} \frac{\sin(2R p_E)}{2R p_E}dR\bigg)
\nonumber \\
&=&\frac{1}{4\pi^2}\frac{2m p_E}{\hbar^2}V\bigg(1 - \frac{\pi}{4p_E}\frac{S}{V}\bigg),
\label{eqA44}
\end{eqnarray}
where we have used that $\int_0^{\infty} \frac{sin(qx)}{x}dx= \frac{\pi}{2}$ if $q>0$.
 
The distribution function (\ref{eqA41}) also allows to obtain the pairing matrix element,
which can be written as 
\begin{widetext}
\begin{equation}
v(E,E') =
\frac{1}{g(E)g(E')}\frac{1}{(2\pi)^6}\int f_E({\bf R},{\bf p}) v({\bf p} - {\bf p'}) f_{E'}({\bf R},{\bf p'}) 
d{\bf p}d{\bf p'}d{\bf R},
\label{eqA45}
\end{equation}
\end{widetext}
where $v({\bf p} - {\bf p'})$ is the Fourier transform of the interaction. In the case of an attractive contact force
$-v_0\delta({\bf R} - {\bf R'})$, it becomes simply $-v_0$ and consequently
\begin{widetext}
\begin{eqnarray}
v(E,E')&=& -\frac{v_0}{g(E)g(E')}\frac{1}{16\pi^4}\left(\frac{2m}{\hbar^2}\right)^2 p_E p_{E'}\int d{\bf R}
\left(1-j_0(2R p_E)\right)\big(1-j_0(2R p_{E'})\big).  
\label{eqA46}
\end{eqnarray}
\end{widetext}
By writing explicitly the spherical Bessel functions and proceeding as before to obtain the density of states (\ref{eqA44}
and taking into account that $\int_0^{\infty} \frac{sin(qx)sin(q'x)}{x^2}dx= \frac{\pi q}{2}$ if $q' \ge q > 0$,
one finally obtains 
\begin{eqnarray}
v(E,E') &=& -\frac{v_0}{g(E)g(E')}\frac{1}{16\pi^4}\left(\frac{2m}{\hbar^2}\right)^2 p_E p_{E'}\\
&\times&\bigg(V - \frac{\pi S}{4p_E} - \frac{\pi S}{4p_{E'}} + \frac{\pi}{4p_E p_{E'}}min(p_E,p_{E'})\bigg), \nonumber
\label{eqA47}
\end{eqnarray}
which in the case of $p_E = p_{E'}$ reduces to
\begin{equation}
v(E,E) = - \frac{v_0}{g(E)^2}\frac{1}{16\pi^4}\left(\frac{2m}{\hbar^2}\right)^2 {p_E}^2 
V\left(1  -\frac{\pi}{4p_{E}}\frac{S}{V}\right). 
\label{eqA48}
\end{equation}  
Now taking into account explicitly the density of states $g(E)$ given by Eq.(\ref{eqA44})
one can write
\begin{equation}
V(E,E) = -\frac{v_0}{V}\left(1 + \frac{\pi}{4p_{E}}\frac{S}{V}\right).
\label{eqA49}
\end{equation}  

In these expressions for the density of states (\ref{eqA44}) and the pairing matrix elements (\ref{eqA49})  
we have assumed that the box is filled by nuclear matter at a given momentum
$p_E$, in spite that the walls of the container compress the matter. To  take into 
account this effect. and following Ref.\cite{Sto-Farine}, we introduce the mass volume $V_M$ and 
surface $S_M$, which are related to the volume and surface of the box by the relationships:
\begin{equation}
V = V_M + \frac{3\pi}{8p_{F}}S_M \qquad S = S_M,
\label{eqA410}
\end{equation}
where $p_{F}$ is the Fermi momentum of the matter in the box but taking into account the
compressional effect. Due to the particle number conservation, it is related to the
Fermi momentum in homogeneous nuclear matter (i.e. in the bulk) $p_{F}^{B}$ by
\begin{equation}  
{p_{F}^3} V_M = {p_{F}^{B}}^3 V  \to  
p_{F} = p_{F}^{B}\left(1 + \frac{\pi}{8p_{F}^{B}}\frac{S_M}{V_M}\right).
\label{A411}
\end{equation}  
Taking into account the compressional effect the Fermi energy of the particles contained the box reads
\begin{equation}   
\mu = \frac{\hbar^2 p_{F}}{2m} = 
\frac{\hbar^2 p_{F}^{B}}{2m}\left(1 + \frac{\pi}{4p_{F}^{B}}\frac{S_M}{V_M}\right)
= \mu^{B}\left(1 + \frac{\pi}{4p_{F}^{B}}\frac{S_M}{V_M}\right).
\label{A412}
\end{equation}
Now we can write the density of states and the matrix in the box at the Fermi energy as
\begin{eqnarray}
g(\mu) &=& \frac{1}{4\pi^2}\frac{2m}{\hbar^2}p_F^B
\left(1 + \frac{\pi}{8p_E^B}\frac{S_M}{V_M}\right) 
\nonumber \\
&\times& V_M \left(1 + \frac{3\pi}{8p_E^B}\frac{S_M}{V_M}\right)
\left(1 - \frac{\pi}{4p_E^B}\frac{S_M}{V_M}\right)\nonumber \\
&=& \frac{1}{4\pi^2}\frac{2m}{\hbar^2}p_F^B V_M\left(1 + \frac{\pi}{4p_E^B}\frac{S_M}{V_M}\right),
\nonumber \\
v(\mu,\mu) &=& - \frac{v_0}{V_M}\left(1 - \frac{\pi}{8p_E^B}\frac{S_M}{V_M}\right). 
\label{A413}
\end{eqnarray}
Using these expressions, we can finally write
\begin{eqnarray}
g(\mu)v(\mu,\mu) &=& -v_0\frac{1}{4\pi^2}\frac{2m}{\hbar^2}p_F^B 
\left(1 + \frac{\pi}{8p_E^B}\frac{S_M}{V_M}\right)\nonumber \\
&=& -v_0 g^B\left(1 + \frac{\pi}{8p_E^B}\frac{S_M}{V_M}\right),
\end{eqnarray}
which is in agreement with the results of Ref.\cite{grain} and where 
$g^B=\frac{1}{4\pi^2}\frac{2m}{\hbar^2}p_F^B$ 
is the density of states per unit of volume in he bulk.

\section{ The pocket formula}
\label{appendix2}
In this Appendix we derive an approximated analytical solution of the gap equation
at the Fermi energy $\mu$. To this end, we start from the TF-BCS gap equation including
subtraction which is necessary for contact interactions with otherwise a diverging gap equation
\begin{widetext}
\begin{equation}
\Delta(E) = - \int dE' g(E') V(E',E) \Delta(E')
\bigg[ \frac{1}{2 \sqrt{(E' - \mu)^2 + \Delta^2(E')}}
- \frac{1}{2(E' - \mu)} \bigg],      
\label{A21}
\end{equation}
\end{widetext}
assuming that the main contribution to the integral in the right hand side of (\ref{A21})
comes from the gaps around the Fermi energy $\mu$, i.e. $\Delta(E') \simeq \Delta(\mu)$, 
we can write this equation for $E=\mu$ as
\begin{equation}
\!\!1\!= \! -\!\! \int \!\!dE' \frac{g(E^\prime) V(E^\prime\!,\mu)}{2}
\bigg[ \frac{1}{\sqrt{(E'\! - \mu)^2\! +\! \Delta^2(\mu)}}
- \frac{1}{(E' \!- \!\mu)} \bigg].
\label{A22}
\end{equation}
With the change of variables $E'=x^2\mu$ and $\Delta(\mu)=\Delta\mu$ ($\Delta$ dimensionless),
we can recast (\ref{A22}) as
\begin{eqnarray}
1&=&\!\!I_1\! +\! I_2\!= \!\!\int^1_0 dx x G(x) 
\bigg[ \frac{1}{\sqrt{(x^2 - 1)^2 + \Delta^2}} + \frac{1}{1-x^2} \bigg]\nonumber\\
&&+ \int^{\infty}_1 dx x G(x)
\bigg[ \frac{1}{\sqrt{(x^2 - 1)^2 + \Delta^2}} - \frac{1}{x^2-1} \bigg],
\label{A23}
\end{eqnarray}
    $G(x)=-g(x^2\mu)V(x^2\mu,\mu)$ with $E'= x^2\mu$. Performing partial integration
and after some algebra the first integrals of (\ref{A23}) reads
\begin{eqnarray}
I_1 &=& \frac{G(x)}{2}\!  \ln\bigg[x^2\!-\!1\! +\! \sqrt{(x^2-1)^2+\Delta^2}\bigg]^1_0 \\
&&- \frac{1}{2} G(x)\left. \ln \frac{\Delta^2}{2(1-x^2)}\right|^1_0 + 2 \int^1_0 dx G(x) \frac{x}{1-x^2}.\nonumber
\label{A24}
\end{eqnarray}
In a similar way, the second integral of (\ref{A23}) reads

\begin{eqnarray}
I_2 \!&=& \!\frac{ G(x)}{2}\bigg[ \ln\big[x^2\!-\!1\! +\! \sqrt{(x^2-1)^2+\Delta^2}\big]
\!-\!\ln(x^2\!-\!1)\bigg]^{\infty}_1\nonumber\\
&&+ \frac{1}{2} \ln 2 G(1),
\label{A25}
\end{eqnarray}

where we have assumed that $G(x)$ vanishes in the limit $x \to \infty$.
Adding (\ref{A24}) and (\ref{A25}) we obtain
\begin{eqnarray}
&&1 = \frac{1}{2} G(x)\ln\bigg[x^2-1 + \sqrt{(x^2-1)^2+\Delta^2}\bigg]^{\infty}_0
\nonumber \\
&-&G(1)\ln \Delta + G(1)\ln 2\nonumber \\ 
&+& \frac{1}{2} G(1) \lim_{x \to \infty}\big[\ln (1-x^2)
+ \ln (x^2-1)\big] \nonumber \\ 
&+&  2 \int^1_0 dx G(x) \frac{x}{1-x^2}.
\label{A26}
\end{eqnarray}
If $G(0)=0$ and we assume that in the limit of $x\rightarrow \infty$ the function $G(x)$ goes to zero faster than $\ln x$ goes to infinity.  
We can write the previous equation as 
\begin{eqnarray}
1&=& -G(1) \ln \Delta + 3 G(1)\ln 2  +  2 \int^1_0 dx G(x) \frac{x}{1-x^2}
\nonumber \\
&+& G(1) \int^1_0 \frac{dx}{1-x^2} - G(1) \int^{\infty}_1 \frac{dx}{x^2-1},
\label{A27}
\end{eqnarray}
where we have used that
\begin{equation}
\int^1_0 \frac{dx}{1-x^2} = \frac{1}{2}\big[\ln 2 - \lim_{x \to 1} \ln(1-x)\big],
\label{A28}
\end{equation}
and
\begin{equation}
\int^{\infty}_1 \frac{dx}{x^2-1} = 
\frac{1}{2}\big[\ln 2 - \lim_{x \to 1} \ln(x-1)\big].
\label{A29}
\end{equation}
Using again Eqs.(\ref{A28}) and (\ref{A29}), we can finally write the gap at the Fermi level as
\begin{equation}
\Delta(\mu)= 8\mu e^{-\frac{1}{G(1)}}
e^{2\int^1_0 dx\frac{x\frac{G(x)}{G(1)}-1}{1-x^2}}.
\label{A210}
\end{equation}
We want now to apply this formula to the specific case of a spherical box. To this end,
we use the density of states (\ref{eqA44}) derived before, which reads
\begin{eqnarray}
g(E)&=& \frac{1}{4\pi^2}\left(\frac{2m}{\hbar^2}\right)^{3/2}\sqrt{E} V - 
\frac{S}{16\pi}\frac{2m}{\hbar^2}\nonumber \\ 
&=&\frac{1}{4\pi^2}\left(\frac{2m}{\hbar^2}\right)^{3/2}\sqrt{E} V\bigg(1 - \frac{\pi}{4 k_E}
\frac{S}{V}\bigg),
\label{A211}
\end{eqnarray}
where we have used that $k_E=\sqrt{\frac{2mE}{\hbar^2}}$.

Using again the change of variables $E=x^2\mu$, we can recast (\ref{A211}) as
\begin{equation}
g(x^2\mu)= 
\frac{1}{4\pi^2}\left(\frac{2m}{\hbar^2}\right)^{3/2}V x\sqrt{\mu} 
\bigg(1 - \frac{\pi}{4 k^B_F}\frac{S}{V}\bigg),
\label{A212}
\end{equation}
where $k^B_F$ is the Fermi momentum in th bulk given by $k^B_F=\sqrt{\frac{2m\mu}{\hbar^2}}$.

However, it is important to point out that the volume $V$ and the surface $S$ correspond to
the borders of the hard sphere. Due to the fact that near the surface the density decreases
 from its bulk value to zero, the relevant mater volume $V_M$, which encloses the right 
number of particles, is smaller than $V$. Thus, taking into account (\ref{eqA410}) we can write
the level density as
\begin{widetext}
\begin{equation}
g(x^2\mu)=
\frac{1}{4\pi^2}\left(\frac{2m}{\hbar^2}\right)^{3/2}x\sqrt{\mu^B}
\bigg(1 + \frac{\pi}{4 k^B_F}\frac{S_M}{V_M}\bigg)
V_M\bigg(1 + \frac{3\pi}{8 k^B_F}\frac{S_M}{V_M}\bigg)
\bigg(1 - \frac{\pi}{4x k^B_F}\frac{S_M}{V_M}\bigg),
\label{A214}
\end{equation}
\end{widetext}
where we have included the correction to the Fermi energy $\mu$ due to the compression effect
of the surface tension (\ref{A412}). 

As we have discussed before, the pairing matrix element corresponding to a $\delta$-force 
around the Fermi energy is given by
\begin{equation}
V(x^2\mu,\mu) = \frac{g_0}{V}\bigg[1 + \frac{\pi}{4}
\frac{min[k(x^2\mu),k(\mu)]}{k(x^2\mu)k(\mu)}\frac{S_M}{V_M}\bigg].
\label{A215}
\end{equation} 
The range of $x$ in the integral of Eq.(\ref{A210}) is from 0 to 1, which implies 
$x<1$ and consequently $V(x^2\mu,\mu)=V(\mu,\mu)$. Therefore, and taking into account 
(\ref{A214}), the integral in the exponent of (\ref{A210}) can be easily computed as
\begin{equation}
\int_0^1 \frac{x\frac{G(x)}{G(1)-1}}{1 - x^2} =
\int_0^1 \frac{x\frac{g(x^2\mu)}{g(\mu)-1}}{1 - x^2} =
-1 -\frac{\pi}{4k^B_F}\frac{S_M}{V_M}.
\label{A216}
\end{equation}
From where follows that
\begin{equation}
e^{2\int^1_0 dx\frac{x\frac{G(x)}{G(1)}-1}{1-x^2}}=
e^{-2}e^{\frac{\pi}{2 k^B_F}\frac{S_M}{V_M}(\ln 2 -1)}.
\label{A217}
\end{equation}
Let us now work out the first factor of the right hand side of Eq.(\ref{A210})
with the help of Eqs.(\ref{A214}) and (\ref{A215})
\begin{equation}
G(1)=-g(\mu) V(\mu,\mu)= -\frac{g_0}{4\pi^2}\frac{2m}{\hbar^2} k^B_F
\bigg(1 + \frac{\pi}{8 k^B_F}\frac{S_M}{V_M}\bigg).
\label{A218}
\end{equation}
From where
\begin{equation}
e^{-\frac{1}{G(1)}}
= e^{\frac{1}{\frac{g_0}{4\pi^2}\frac{2m}{\hbar^2}k^B_F}}
e^{-\frac{1}{\frac{g_0}{4\pi^2}\frac{2m}{\hbar^2}k^B_F\frac{\pi}{8k^B_F}\frac{S_M}{V_M}}}.
\label{A219}
\end{equation}
Combining Eqs.(\ref{A217}) and (\ref{A219}), taking into account the correction to the Fermi
energy due to the compression and the gap in the bulk given by \cite{Grasso}
\begin{equation}
\Delta_B = 8\mu_B e^{\frac{1}{\frac{g_0}{4\pi^2}\frac{2m}{\hbar^2}k^B_F}-2},
\label{A220}
\end{equation}
one recovers Eq.(14) of the main text. 
\begin{equation}
  \Delta(\mu) = \Delta_B\bigg (1+\frac{\pi}{4k^B_F}\bigg )e^{-\frac{4\pi^2\hbar^2}
{g_0 2mk^B_F}\frac{\pi}{8k^B_F}\frac{S_M}{V_M}}e^{\frac{\pi}{2k^B_F}
\frac{S_M}{V_M}(\ln 2 - 1)}.
\label{A221}
 \end{equation}
It may also be useful to give the pocket formula for finite range forces as, e.g., the Gogny D1S force used in this paper. 
It can be derived in analogy to the case with delta force and subtraction presented just above

\begin{equation}
  \Delta(\mu) = 8\mu e^{-\frac{1}{g(\mu)V(\mu,\mu)}} e^R.
    \label{PF-F}
\end{equation}
In the above formula, one should count the chemical potential $\mu$ from the bottom of the mean field potential. In weak coupling one then has to good approximation $\mu \sim \varepsilon_F$, the Fermi energy. For the factor $R$ one gets

\begin{equation}
  R= \int_0^{\infty} dx\frac{x\frac{G(x)}{G(1)}-1}{|1-x^2|},
  \label{para-R}
\end{equation}
with
\[G(x) = -g(x^2\mu)V(x^2\mu,\mu). \]
Supposing a local mean field, it may be useful to split the $R$-factor into three pieces: $R=R_1+R_2+R_3$ with

\begin{equation}
  R_1 = \int_0^1  dx\frac{x\frac{G(x)}{G(1)}-1}{1-x^2},
  \label{R1}
\end{equation}

\begin{equation}
  R_2 = -\int_1^{\sqrt{|U_0|/\mu}}  dx\frac{x\frac{G(x)}{G(1)}-1}{1-x^2},
  \label{R2}
\end{equation}
where $|U_0|$ is the upper border of the mean field counted from its bottom.
\begin{equation}
  R_3 = -\int_{\sqrt{|U_0|/\mu}}^{\infty} dx \frac{1}{x^2-1}.  
    \label{R3}
\end{equation}
This form of $R_3$ holds because for $E \rightarrow |U_0|$, the matrix
element $V(E,\mu)$ and therefore $G(x)$ tends to zero. The integral in $R_3$ can be performed to obtain

\begin{equation}
  e^{R_3} = \sqrt{\frac{ \sqrt{|U_0|/\mu}-1}{\sqrt{|U_0|/\mu}+1}}.
  \label{R3-a}
\end{equation}
Above formula for $R_3$ is interesting, since it immediately tells us that the gap will break down towards the drip line.\\

The coefficients $R_1$ and $R_2$ can further be approximated for systems with ingredients similar to nuclear mean fields and effective pairing forces for which it turns out that

\[ \frac{G(x)}{G(1)} \simeq x \]
With this approximation this then leads to Eq.(\ref{PF-F1}) of the main text.

To apply the pocket formula (\ref{PF-F}) in the case of a finite-range pairing interaction, an essential ingredient is the
pairing matrix element $V(E,E')$, which in the case of the Gogny force is given by
\begin{widetext}
\begin{equation}
V(E,E') = \frac{1}{4 \pi^3 g(E) g(E')}\left(\frac{2m}{\hbar^2}\right)^2 \sum_{i=1}^{i=2} \frac{z_{c_i}}{\mu_i}
\int_0^{R_{min}} dR R^2 e^{-\frac{\mu_i^2(k_E + k_{E'})}{4}} \sinh \left(\frac{\mu_i^2 k_E k_{E'}}{2}\right),
\label{meg}
\end{equation}
\end{widetext}
where $z_{c_i}= 2 \pi^{3/2}(W_i-B_i-H_i+M_i)$ are the stregths of pairing interaction, $k_E$ and  $k_{E'} = k(E',R)$
the local Fermi momenta at energies $E$ and $E'$ with the mean-field
potential $U(R)$. $g(E)$ and $g(E')$ are
the corresponding DOS at these energies. The upper limit in the integral is $R_{min}=min(R^T_E,R^T_{E'})$,
which is  the smallest of the classical turning points in the local Fermi momenta.

\section{ Quantal study of a free particle in a hemi-sphere}
\label{appendix3}

The Schr\"odinger equation for a free particle confined in a half-sphere of radius $R$ read

\begin{eqnarray}\label{sh:half}
-\frac{\hbar^2}{2m}\nabla^2\psi_{nlm}(r,\theta,\phi) = E\psi_{nlm}(r,\theta,\phi).
\end{eqnarray}

The problem has  very specific boundary conditions $i.e.$ $\psi_{nlm}(r,\theta,\phi)$ is equal to zero at the edges 
of the half sphere. Before entering in a more detailed discussion, it is interesting to specify the system of reference.
$r$ represent the radial distance from the centre of the sphere and it can take all values in the range $r\in[0,R]$. $\theta$ 
is the angle formed by $r$ with the axis of symmetry of the half sphere it goes from $\theta\in[0,\pi/2]$.
The value $\theta=0$  indicates a point on the top of the half-sphere, the value $\theta=\pi/2$ is placed on the disk 
closing the bottom of the half-sphere.
$\phi$ is the angle formed by the projection of r on $xy$-plane and it spans from 0 to $2\pi$.

The boundary conditions can be then translated as

\begin{eqnarray*}
\begin{array}{cc}
\psi_{nlm}(R,\theta,\phi)=0 &\theta\ne \frac{\pi}{2},\\
\psi_{nlm}(r,\frac{\pi}{2},\phi)=0 &\theta= \frac{\pi}{2},
\end{array}
\end{eqnarray*}

these two conditions take into account the fact that the wave-function is zero on the hemisphere and the disk delimiting the Hilbert space.

To solve such a problem, we use the usual ansatz of
\begin{eqnarray}
\psi_{nlm}(r,\theta,\phi)\propto u_{nl}(r)Y_{lm}(\theta,\phi),
\end{eqnarray}

\noindent here $nlm$ are the quantum number of the system. $Y_{lm}(\theta,\phi)$ is the spherical harmonic. Notice we neglect spin terms since there is no spin-orbit 
coupling and the terms simply factorise.
To prove that this is an eigenvector, we need to check that once replaced into the original differential equation we get an identity.

We thus write the laplacian in spherical coordinates

\begin{equation*}
\Delta= \frac{1}{r^2}\frac{\partial}{\partial r}\left(r^2\frac{\partial }{\partial r}\right) + \frac{1}{r^2\sin\theta}\frac{\partial}{\partial\theta}\left(\sin\theta\frac{\partial }{\partial\theta}\right) + \frac{1}{r^2\sin^2\theta}\frac{\partial^2}{\partial\phi^2}. 
\end{equation*}

We replace in Eq.\ref{sh:half} and we get

\begin{eqnarray}
-\frac{\hbar^2}{2m}\left[ \frac{1}{r^2}\frac{\partial}{\partial r}\left(r^2\frac{\partial }
{\partial r}\right)- \frac{l(l+1)}{r^2}\right]u_{nl}(r) =E_{nl}u_{nl}(r).\nonumber \\
\end{eqnarray}

By defining  $k_{nl}^2=\frac{2m}{\hbar^2} E_{nl}$ and performing simple manipulations, we obtain  the Bessel equation whose solution are the spherical Bessel functions $u_{nl}(r)=j_{l}\left( k_{nl}r \right)$.

By imposing the first set of boundary conditions, i.e. $u_{nl}(R)=0$ we obtain the discretisation of the eigen-spectrum. This condition is angle independent. In the case of a closed half-sphere, we also need to impose that the wave function is zero on the bottom disk closing the space.
This condition should be valid for any value of $r$ and fixed $\theta=\pi/2$.

As such we have to impose this on the angular part of the wave function

\begin{equation}
Y_{lm}\left(\frac{\pi}{2},\phi\right)=0.
\end{equation}

This is achieved simply by selecting the associated Legendre polynomial that are odd $P_l^m(\cos\theta) $, or in other words they have a node on the $\theta=\pi/2$ plane.
This conditions is respected for all polynomials so that $l+m$ is an odd number ($l=0$ excluded).

As a consequence, we can solve the half-sphere problem using the same methodology used to solve the sphere, apart from the extra selection rule on the quantum number used to build the basis.

  \end{appendix}


\begin{thebibliography}{0}
\expandafter\ifx\csname natexlab\endcsname\relax\def\natexlab#1{#1}\fi
\expandafter\ifx\csname bibnamefont\endcsname\relax
  \def\bibnamefont#1{#1}\fi
\expandafter\ifx\csname bibfnamefont\endcsname\relax
  \def\bibfnamefont#1{#1}\fi
\expandafter\ifx\csname citenamefont\endcsname\relax
  \def\citenamefont#1{#1}\fi
\expandafter\ifx\csname url\endcsname\relax
  \def\url#1{\texttt{#1}}\fi
\expandafter\ifx\csname urlprefix\endcsname\relax\def\urlprefix{URL }\fi
\providecommand{\bibinfo}[2]{#2}
\providecommand{\eprint}[2][]{\url{#2}}

\end{thebibliography}


\begin{thebibliography}{99}
  \bibitem{RS}
    P. Ring, P. Schuck, {\it The nuclear Many-Body Problem}, Springer 1980.
  \bibitem{BM}
    A. Bohr, B. R. Mottelson, {\it Nuclear Structure}, Vol.I, Benjamin, New York 1975.
  \bibitem{Anderson}
    P. W. Anderson, J. Phys. Chem. Solids {\bf 11}, 26 (1959).
  \bibitem{Shanenko06}
    A. A. Shanenko, M. D. Croitoru, Phys. Rev. B{\bf 73}, 012510 (2006).
  \bibitem{Peeters}
    A. A. Shanenko, M. D. Croitoru, M. Zgirsky, F. M. Peeters, K. Arutyunov, Phys. Rev. B {\bf 74}, 052502 (2006).
  \bibitem{Shanenko07}
    A. A. Shanenko, M. D. Croitoru, F. M. Peeters, Phys. Rev. B {\bf 75}, 014519 (2007).
  \bibitem{Shanenko09}
    Yajiang Chen, M. D. Croitoru, A. A. Shanenko, F. M. Peeters, J. Phys.: Condensed Matter, {\bf 21}, 435701 (2009).
  \bibitem{Shanenko11}
    M. D. Croitoru, A. A. Shanenko, C. C. Kaun, F. M. Peeters, Phys. Rev. B {\bf 83}, 214509 (2011).

  \bibitem{G-G}
    J. Mayoh, A. M. Garcia-Garcia, arXiv:1309.6255.

  \bibitem{GG1}
    A. M. Garcia-Garcia, J. D. Urbina, E. A. Yuzbashyan, K. Richter, B. L. Altshuler, Phys. Rev. B {\bf 83}, 014510 (2011).

  \bibitem{GG2}
    A. M. Garcia-Garcia, Phys. Rev. B 90,014509 (2014).

    
  \bibitem{Nature}
    S. Bose, A. M. Garcia-Garcia, M. M. Ugeda, J. D. Urbina, C. H. Michaelis,
   I. Brihuega, K. Kern, Nature Materials 9, 550 (2010). 

 \bibitem{50yearsBCS}
   P. Schuck, X. Vi\~nas
   {\it Fifty Years of Nuclear BCS}, Edited by R.A. Broglia and V. Zelevinsky,
   World Scientific Publishing Co Pte Ltd, Singapore, 212 (2013).
   

   
  \bibitem{grain}
    M. Farine, F. W. J. Hekking, P. Schuck, X. Vi\~nas, Phys. Rev. B {\bf 68}, 024507 (2003).

  \bibitem{Faessler}
    P. Schuck, R. W. Hasse, J. Jaenicke, C. Gr\'egoire, B. R\'emaud, F. S\'ebille, E. Suraud,  Progr. Part. Nucl. Phys. 22, 181 (1989).

  \bibitem{Annals}
    M. Cerntelles, P. Schuck, X. Vi\~nas, Ann. Phys. 322, 489 (2007).
    
    
  \bibitem{Grasso}
   M. Grasso and M. Urban,
Phys. Rev. {\bf A68}, 033610 (2003).
  \bibitem{Bal-Bloch}
    R. Balian, C. Bloch, Ann. Phys. (N.Y) {\bf 60}, 401 (1970).
  \bibitem{Ale}
    A. Pastore, P. Schuck, M. Urban, X. Vi\~nas, J. Margueron, Phys. Rev. A {\bf 90}, 043634 (2014).
  
      
  \bibitem{Sto-Farine}
    W. Stocker, M. Farine, Ann. Phys. (N.Y.) {\bf 159}, 255 (1985).
  \bibitem{Abramo-Steg}
   M. Abramowitz and I.A. Stegun,
{\it Handbook of mathematical Functions}
(Dover, New York, 1972) ch.17.
  \bibitem{Bhaduri-Brack}
    M. Brack, R. J. Bhaduri, {\it Semiclassical Physics}, CRC Press, Taylor and Francis group. 
  \bibitem{Hasse}
    R. W. Hasse, Nucl. Phys. {\bf A467}, 407 (1987).
  \bibitem{Gogny}
    J. F. Berger, M. Girod, D. Gogny, Computer Phys. Communications {\bf 63}, 365 (1991). 
  \bibitem{bertsch91}
    G. F. Bertsch, H. Ebensen,
   Ann. Phys. (N.Y) {\bf 209}, 327 (1991).    
  \bibitem{garrido99}
   E. Garrido, P. Sarriguren, E. Moya de Guerra, P. Schuck,
   Phys. Rev. C {\bf 60}, 064312 (1999). 
  \bibitem{tian09}Y. Tian Z.Y. Ma, P. Ring, 
   Phys. Lett. B {\bf 676}, 44 (2009).
 \bibitem{shlomo91}
    S. Shlomo,  Nucl. Phys. {\bf A539}, 17 (1991).

  \bibitem{Kohn}
    N. D. Lang, W. Kohn, Phys. Rev. B, 4555 (1970).

  \bibitem{sc-me}
    X. Vi\~nas, P. Schuck, M. Farine, M. Centelles, Phys. Rev. C {\bf 67}, 05430\
    7 (2003).
    


    
  \bibitem{Bulgac}
    A. Bulgac, Y. Yu, Phys. Rev. Lett. {\bf 88}, 042504 (2002).
  \bibitem{me-paper}
    X. Vi\~nas, P. Schuck, M. Farine, M. Centelles, Phys. Rev. C {\bf 67}, 014324 (2003).

  \bibitem{stability}
    P. Marmier and E. Sheldon, {\it Physics of Nuclei and Particles} (Academic Press, London, 1969).

    
  \bibitem{HFB}
    HFB, ...
  \bibitem{RingBCS}
    J. Xiang, Z. P. Li, J. M. Jao, W. H. Long, P. Ring, J. Meng, Phys. Rev. C {\bf 88}, 057301 (2013).
  \bibitem{soubbotin00} 
  V.B. Soubbotin, X. Vi\~nas, Nucl. Phys. {\bf A665}, 2912 (2000).
  \bibitem{soubbotin03}
  S. Krewald, V.B. Soubbotin, V.I. Tselyaev, X. Vi\~nas,  Phys. Rev. C {\bf 74}, 064310 (2006). 
  \bibitem{krewald06}
   V.B. Soubbotin, V.I. Tselyaev, X. Vi\~nas,  Phys. Rev. C {\bf 67}, 054307 (2003).
  \bibitem{IJMP2011}
    X. Vi\~nas, P. Schuck, M. Farine, Int. J. Mod. Phys. E Vol.20, No. 2, 399 (2011).

  \bibitem{pastore13}
   A. Pastore, J. Margueron, P. Schuck, X. Vi\~nas
   Phys. Rev. C {\bf 88}, 034314 (2013)

   



\end{thebibliography}
\end{document}